\documentclass[journal]{IEEEtran}

\usepackage{cite}

\ifCLASSINFOpdf
  \usepackage[pdftex]{graphicx}
  \graphicspath{{../pdf/}{../jpeg/}{./figures/}{./figures/experiments/}{./figures/svg}}
  \DeclareGraphicsExtensions{.pdf, .jpeg, .jpg, .png, .eps}
\else
\fi
\usepackage{amsmath}
\usepackage{empheq}

\usepackage{algorithmic}

\newcounter{myalgorithm}
\newcounter{PlaceHolder}

\newenvironment{myalgorithm}
{\setcounter{PlaceHolder}{\value{figure}} 
 \setcounter{figure}{\value{myalgorithm}}\begin{figure} 
\renewcommand{\figurename}{Alg.} }
{\end{figure} \setcounter{figure}   
{\value{PlaceHolder}} \addtocounter {myalgorithm} {1}}

\ifCLASSOPTIONcompsoc
  \usepackage[caption=false,font=normalsize,labelfont=sf,textfont=sf]{subfig}
\else
  \usepackage[caption=false,font=footnotesize]{subfig}
\fi

\usepackage{amssymb}
\usepackage[capitalise]{cleveref}
\usepackage{multirow}

\DeclareSymbolFont{bbold}{U}{bbold}{m}{n}
\DeclareSymbolFontAlphabet{\mathbbold}{bbold}

\usepackage{pgfplots}
\usepackage{lipsum,adjustbox}
\pgfplotsset{compat=newest}
\pgfplotsset{plot coordinates/math parser=false}
\newlength\figureheight
\newlength\figurewidth 

\usepackage{amsmath,amsfonts,amssymb,rotating}

\mathcode`l="8000
\begingroup
\makeatletter
\lccode`\~=`\l
\DeclareMathSymbol{\lsb@l}{\mathalpha}{letters}{`l}
\lowercase{\gdef~{\ifnum\the\mathgroup=\m@ne \ell \else \lsb@l \fi}}%
\endgroup

\newfont{\bbb}{msbm10 scaled 500}

\newfont{\bb}{msbm10 scaled 1100}

\newcommand{\Ac}{{\cal A}}

\newcommand{\Cc}{{\cal C}}

\newcommand{\Rc}{{\cal R}}

\renewcommand{\l}{(l)}

\newcommand{\<}{\left\langle}
\renewcommand{\>}{\right\rangle}

\def\<{\langle}
\def\>{\rangle}

\newtheorem{theorem}{Theorem}

\newtheorem{constraint}[theorem]{Constraint}

\usepackage{mathtools,cases}

\begin{document}
%
\title{Fine granularity access in interactive compression of 360-degree images based on rate-adaptive channel codes}
\author{Navid~Mahmoudian~Bidgoli,
        ~Thomas~Maugey,
        ~Aline~Roumy
\thanks{The authors are with the Inria, Univ Rennes, CNRS, IRISA (e-mail: \textit{\{navid.mahmoudian-bidgoli, thomas.maugey, aline.roumy\}@inria.fr}).}
\thanks{This work was supported by the Cominlabs excellence laboratory with funding from the French National Research Agency (ANR-10-LABX-07-01) and by the Brittany Region (Grant No. ARED 9582 InterCOR).}}


\maketitle

\vspace*{-7cm}
\fbox{%

    \parbox{0.9\textwidth}{%
        This is a pre-print of an article published in IEEE Transactions on Multimedia. The final authenticated version is available online at: https://doi.org/10.1109/TMM.2020.3017890
    }%
}%
\vspace*{5.75 cm}

\begin{abstract} 
In this paper, we propose a new interactive compression scheme for omnidirectional images. This requires two characteristics: efficient compression of data, to lower the storage cost, and random access ability to extract part of the compressed stream requested by the user (for reducing the transmission rate). For efficient compression, data needs to be predicted by a series of references that have been pre-defined and compressed. This contrasts with the spirit of random accessibility. We propose a solution for this problem based on incremental codes implemented by rate-adaptive channel codes. This scheme encodes the image while adapting to any user request and leads to an efficient coding that is flexible in extracting data depending on the available information at the decoder. Therefore, only the information that is needed to be displayed at the user's side is transmitted during the user's request, as if the request was already known at the encoder. The experimental results demonstrate that our coder obtains a better transmission rate than the state-of-the-art tile-based methods at a small cost in storage. Moreover, the transmission rate grows gradually with the size of the request and avoids a staircase effect, which shows the perfect suitability of our coder for interactive transmission.

\end{abstract}


%
\IEEEpeerreviewmaketitle

%

\section{Introduction}
Omnidirectional or 360$^{\circ}$ images/videos are becoming increasingly popular as they offer an  immersive experience in virtual environments,  and are available with cheap devices, \emph{i.e.}, omnidirectional cameras and Head-Mounted-Displays (HMD).
As a result, a gigantic number of 360$^{\circ}$ videos has been uploaded on video streaming platforms. 
The specificity of 360$^{\circ}$ contents is that they are accessed in an \emph{interactive} way,  \textit{i.e.} 
   only a small portion, called \emph{viewport},  of the whole image is requested by the user.
This has two consequences. First, to reach a sufficient resolution at the observer's side, the global resolution of the input signal is significantly high, (typically 4K or 8K). This underlines the need for \emph{efficient compression} algorithms. Second, to save bandwidth, the coding algorithm should allow \emph{random access} to the compressed data, and only transmit what is requested.
However, efficient compression introduces dependencies in the compressed representation, which is in contrast to the need of random accessibility.


\begin{figure}[h]
\centering
\includegraphics[width=0.6\linewidth]{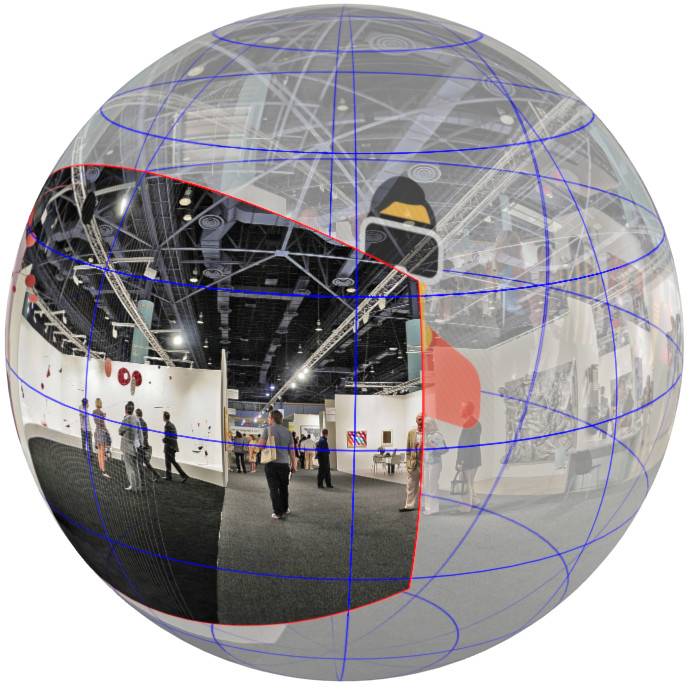}
\caption{At a given instant, a user only observes a small portion of the 360$^\circ$ image. The goal of interactive coding schemes is to transmit only this useful information while keeping a good global compression efficiency.}
\label{fig:user_viewport}
\end{figure}

Different approaches have been proposed to solve this tradeoff in the context of  360$^{\circ}$ data. All these solutions rely on powerful 2D video compression tools such as Versatile Video Coding (VVC), and thus on the projection of the acquired spherical content onto a 2D plane (equirectangular \cite{snyder1997flattening}, cube map \cite{1377362,kuzyakov2016next}, dodecahedron \cite{4895316}, pyramid \cite{kuzyakov2016next}). 
A first compromise solution consists in sending the entire 360$^\circ$ image, achieving a high compression efficiency but a low  random accessibility.
A second approach consists in partitioning the 2D projected 360$^{\circ}$ image into sub-parts, called \emph{tiles}, and encoding them separately  \cite{Zare:2016:HTS:2964284.2967292,Qian:2016:OVD:2980055.2980056}.  The transmission rate is significantly reduced because only the  tiles necessary to reconstruct the requested viewport are transmitted. However, the approach has two main drawbacks: first, the correlation between the tiles is not exploited, and second, the sent tiles contain usually significantly more data than the requested viewport. Therefore, the tile-based approach also fails to achieve the transmission rate of an \emph{oracle coder} that knows the request in advance and encodes the requested viewport(s) only. To achieve this ideal transmission rate, one could encode and store all possible requested viewports, but this exhaustive solution leads to a huge storage cost.

\begin{figure*}
\includegraphics[width=1\linewidth]{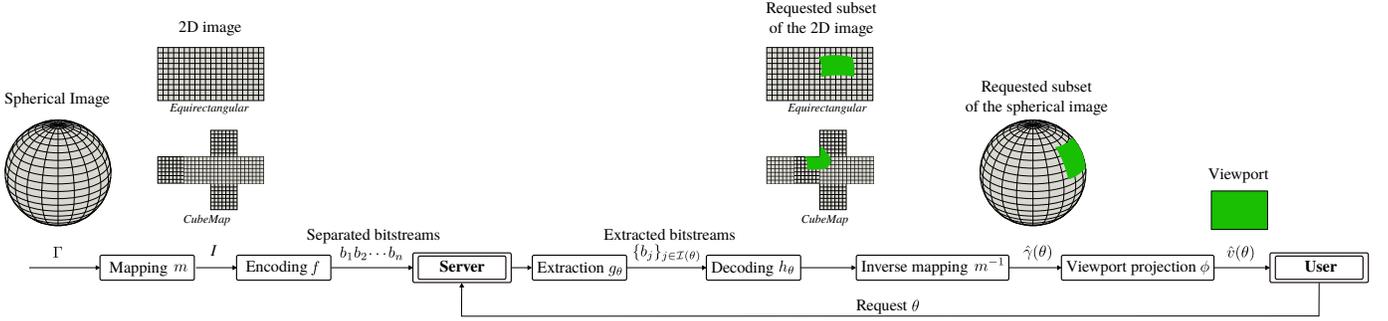}
\caption{Interactive compression of a spherical image $\Gamma$. Mapping and encoding operations are performed offline whereas extraction, decoding and viewport construction are performed online, upon user's request.}
\label{fig:interactive_compression_360}
\end{figure*}

In this paper, we propose a novel 360$^\circ$ image interactive coding scheme that achieves the oracle transmission rate, while keeping the storage cost reasonable. More precisely, the storage size is far less than the exhaustive storage solution and close to the solution without random access (compression of the entire 360$^\circ$ image).

Classical compression schemes work block-wise, and use  a fixed block scanning order for compression and decompression. 
Therefore, in 360$^\circ$ tile-based image compression, the smallest entity that can be accessed is a tile, a predefined set of blocks.
By contrast, we propose to encode the blocks such that \emph{any decoding order} is possible at the decoder's side. As a consequence, i) only the useful blocks are sent and decoded,  ii) the correlation between the blocks is still taken into account. Therefore, i) and ii) enable our coder to reach the same transmission rate as the oracle coder.
In practice, for each block to be encoded, several predictions are generated (corresponding to different block decoding orders), and a bitstream is built based on rate-adaptive channel codes studied in  \cite{7351125, dupraz2019rate, maugey2020incremental}. This bitstream compensates for the worst prediction, and a substream can be extracted from it to correct any other predictions.

Moreover, to build an efficient end-to-end coder, we also propose novel prediction modes to adapt not only to interactivity, but also to the geometrical properties and discontinuities of the  360$^{\circ}$ mapped data.
Then, a new strategy is developed to spread the \emph{access points} (independently standalone decodable blocks) such that the decoding process can be triggered whatever the requested viewport is. Then, a new scanning order is proposed in order to further lower the transmission rate.
The experimental results demonstrate the ability of our proposed coder to reach the same transmission rate as the oracle, while incurring a reasonable additional storage cost.

The following of this paper is organized as follows. The compression of 360$^\circ$ content with the user-dependent transmission is formulated in \cref{sec:problem_formulation}. The overview of the proposed coder is given in \cref{sec:overview_interactive_coding}. In \cref{sec:details} the details of the proposed coder are explained. 
Finally, in \cref{sec:experiments} we compare our approach with baseline coders, including the tiling approach.


%

\section{Problem formulation for Interactive compression of $360^{\circ}$ content} \label{sec:problem_formulation}

Before describing our core contributions on interactive compression of 360$^\circ$ images (in Sec. \ref{sec:overview_interactive_coding} and  \ref{sec:details}),  we first overview the interactive 360$^\circ$ coding model. This model, depicted on \cref{fig:interactive_compression_360},  is general and encompasses state-of-the-art but also the proposed scheme. Then, the coding scheme optimization problem is formulated and different measures for evaluation are proposed.

\subsection{Data model, projection and compression for storage on the server }
A  360$^\circ$ data, denoted by $\Gamma$, is a \emph{spherical image},  \textit{i.e.}, a set of pixels defined on the sphere. Each pixel is determined by its value (\textit{e.g.}, R,G,B) and its spherical coordinates $\theta$,  \textit{i.e.}, a 2D vector composed of the latitude  $\in[-\pi/2,\pi/2]$, and the longitude  $\in[-\pi,\pi)$. To be able to use classical compression algorithms, this data $\Gamma$ is first mapped onto a Euclidean space. This leads to $I$, the Euclidean 2D representation, also called the \emph{2D image}. 
For instance, the equirectangular projection consists in uniformly sampling the spherical coordinates $\theta$, and in replacing $\theta$ by Euclidean coordinates: $(x,y)\approx\theta$.
The mapping function from the sphere to the plane is denoted by
$$m: \Gamma \mapsto I.$$

Then, compression is performed with the encoding function $f$. To allow interactivity, the compression leads to a set of independently extractable bitstreams $(b_1,\ldots,b_n)$:
\begin{align} \label{eq:separate_bitstream}
  (f\circ m) (\Gamma) =  (b_1,\ldots,b_n), 
\end{align}
where $\circ$ stands for the composition of functions.  The \emph{storage cost}, \textit{i.e.}, the size of the data stored on the server is defined as:
\begin{equation} \label{eq:storage}
S = |f \circ m (\Gamma)| = \sum_{i=1}^n | b_i |.
\end{equation}

\subsection{Bitstream extraction and decoding upon user's request}
In interactive compression, the viewer points in the direction $\theta$ to get a subset $\gamma(\theta)$ of the whole spherical content $\Gamma$.
Upon request, the server extracts the bitstreams with indices $i\in\mathcal{I}_{\theta }$
$$ g_{\theta}: (b_1,\ldots,b_n) \mapsto (b_{i})_{i\in\mathcal{I}_{\theta }}, $$
and sends it to the user at rate
\begin{equation}\label{eq:transmission_rate}
R_\theta = | g_\theta(b_1,\ldots,b_n) | = \sum_{i \in \mathcal{I}(\theta)} |b_i|,
\end{equation}
where $R_\theta$ is  called the \emph{transmission rate}.

At the user's side, the extracted bitstreams allow to reconstruct ${\gamma}(\theta)$, \textit{i.e.}, the requested subset of the spherical content:
$$ m^{-1}\circ h_\theta:  (b_{i})_{i\in\mathcal{I}_{\theta}} \mapsto \hat{\gamma}(\theta), $$
where $h_\theta$ is the decoding function of  the received bitstreams. Note that due to lossy compression  $\gamma(\theta) \neq \hat{\gamma}(\theta)$.

Finally, a 2D image, called \emph{viewport} $v(\theta)$ is displayed to the user  (\cref{fig:interactive_compression_360}).  This image is the projection (denoted $\phi$) of $\gamma(\theta)$, the requested portion of the spherical image, onto a plane orthogonal to the direction $\theta$, and tangent to the sphere. Note that the size and resolution of the viewport determines the subset  $\gamma(\theta)$ of pixels requested.

Evaluating the quality of 360-degree images is a wide problem that has been intensively studied by the community. A comprehensive review and analysis is given in \cite{8756213}. To measure a realistic quality of experience at the user side, the distortion is measured on the image displayed to the user \cite{7328056}. Formally, \emph{the distortion} is
\begin{equation}\label{eq:distortion}
D_\theta = || v(\theta) - \hat{v}(\theta) ||_2^2 = || \phi(\gamma(\theta)) - \phi(\hat{\gamma}(\theta)) ||_2^2.
\end{equation}

%
%
\subsection{Storage-Transmission trade-off in state-of-the-art methods}	\label{subsec:state_of_the_art}

State-of-the-art 360$^\circ$ coders achieve interactivity, or equivalently separable bitstreams \eqref{eq:separate_bitstream}, by dividing the  
2D image $I$ into sub-elements and encode each sub-element independently with $f$,  the encoding function of a classical 2D video codec, such as VVC. For instance, in the case of the cubemap projection, the sphere is projected on a cube and each sub-element is a face of the cube  \cite{kuzyakov2016next, 1377362}. Similarly,  \cite{7823595} divides the sphere into $6$ elements.
In the case of the equirectangular projection \cite{Zare:2016:HTS:2964284.2967292,Qian:2016:OVD:2980055.2980056,8122230,7996611}, the 
2D image is partitioned into sub-images called \emph{tiles}, and each tile is coded separately.

Thus, all state-of-the-art approaches use the same encoding function $f$, and they only differ in the mapping $m$. So, these methods attempt to solve
\begin{equation} \label{eq:conventional_interactive_cost_function}
\begin{aligned}
\min_{m} \lambda S+\mathbb{E}_{\theta}(R_\theta), \\
\text{such that } \mathbb{E}_{\theta}(D_\theta) \leq \delta.
\end{aligned}
\end{equation}
by proposing different mapping functions $m$, where $\mathbb{E}_{\theta}$ is the expectation over all users' requests.

\subsection{Our goal: oracle transmission rate with small storage cost}
%


Our coder differs from state-of-the-art methods in several ways. First, our goal is to achieve the same transmission rate $\mathcal{R^*}$ as the oracle scheme that knows the request in advance (\textit{i.e.}, upon encoding). Second, the storage should not be significantly penalized. Finally, to achieve this goal, the encoding and decoding functions $(f,g_\theta,h_\theta)$ are modified, but not the mapping $m$. So, our method attempts to solve
\begin{equation} \label{eq:conventional_interactive_cost_function}
\begin{aligned}
\min_{f, g_\theta,h_\theta} S, \\
\text{such that } & \mathbb{E}_{\theta}(D_\theta) \leq \delta, \\
& \mathbb{E}_{\theta}(R_\theta) = \mathcal{R^*},
\end{aligned}
\end{equation}
by proposing new functions $(f,g_\theta,h_\theta)$. 
Since the proposed method is compatible with any mapping $m$, we adopt, without loss of generality, the equirectangular projection in the rest of this paper.
%

%

%

\section{Principle and overview of the proposed interactive coding scheme} \label{sec:overview_interactive_coding}
\begin{figure}[h]
\centering
\includegraphics[width=.7\linewidth]{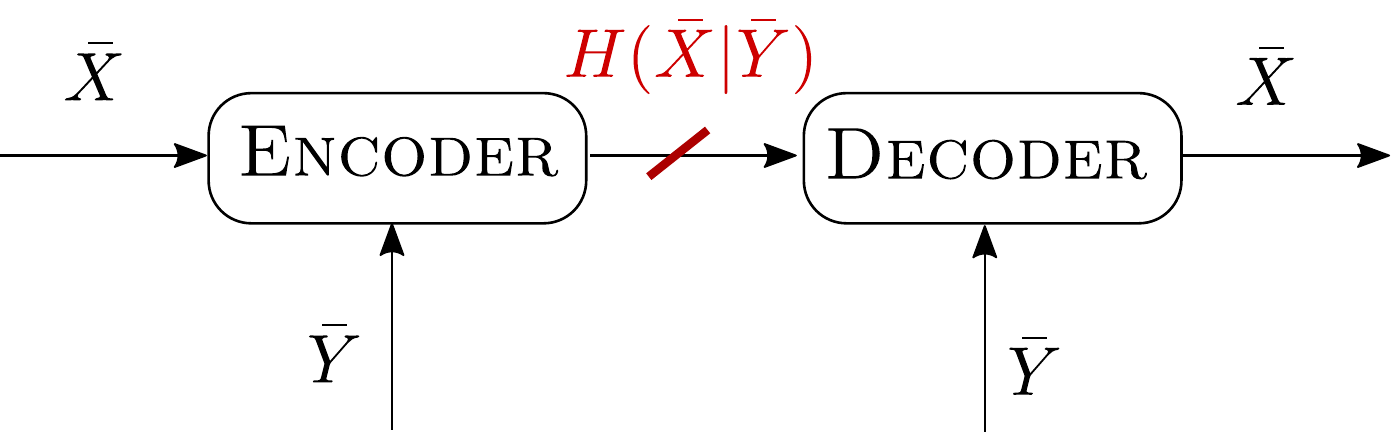}
\caption{Lossless conditional source coding \cite{cover2006b}:  when SI $\bar{Y}$ is available at both encoder and decoder, the source $\bar{X}$ is compressed at rate $H(\bar{X}|\bar{Y})$, which is smaller than its entropy $H(\bar{X})$. 
}
\label{fig:Ccoding}
\end{figure}

The strength of conventional coders resides in the block-wise compression of the image, where the correlation between the blocks is captured thanks to a \emph{predictive coding} scheme \cite{wien2014high}. Predictive coding is based on the principle of \emph{conditional coding}, illustrated in \cref {fig:Ccoding}.
 In this scheme, an i.i.d. source $\bar{X}$  is compressed losslessly, while another i.i.d. source $\bar{Y}$, called \emph{side information} (SI), is available at both encoder and decoder. 
The achievable compression rate is the conditional entropy $H(\bar{X}|\bar{Y})$ \cite{cover2006b}, which is smaller than the rate $H(\bar{X})$ achieved when no SI is available. This shows the benefit of the SI. %
  $\bar{X}$ models the image block to be compressed, and $\bar{Y}$ models the prediction of $\bar{X}$ using another block.

%

\begin{figure}[h]
\centering
\includegraphics[width=1\linewidth]{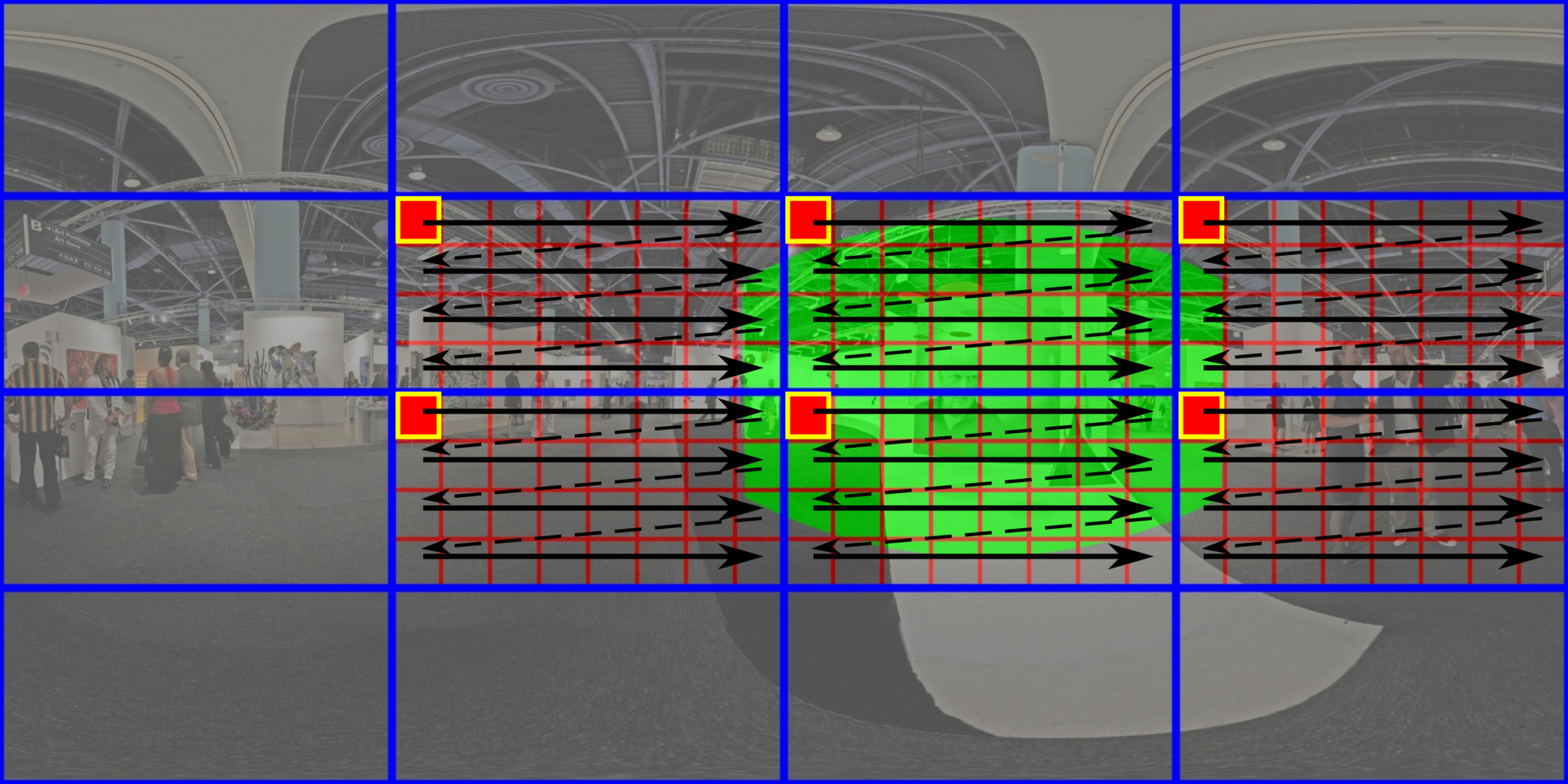}
\centering{(a) tile-based approach}\vspace{0.3cm}
\includegraphics[width=1\linewidth]{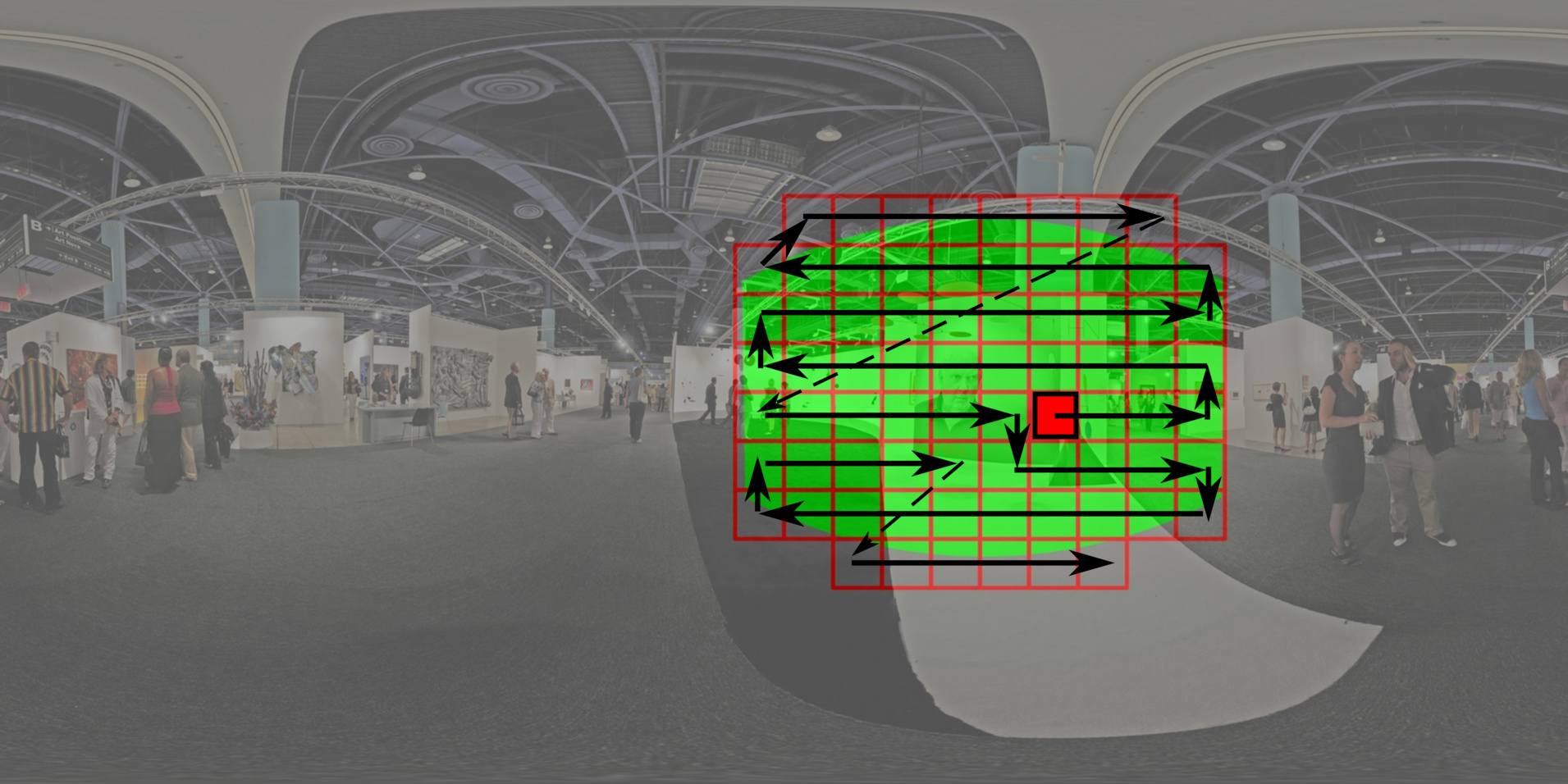}
\centering{(b) our coding scheme}
\caption{
The green region is the requested portion of the 2D image. The transmitted blocks are represented with rectangles with red borders (including the filled ones that correspond to the reference or access blocks), and the arrows depict the block decoding order. For the same request, the proposed scheme (b) reduces the amount of information sent but not displayed compared to the tile-based approach (a). Moreover, the correlation among the blocks is still taken into account with our coder, while the tile-based approach compresses each tile separately.}
\label{fig:tiles_and_blocks}
\end{figure}

This scheme is efficient but imposes a fixed pre-defined encoding and decoding order of blocks, which contradicts the idea of randomly accessing any part of an image. Therefore, to allow random access, state-of-the-art 360$^\circ$ coders (see \cref{subsec:state_of_the_art}) partition the image into subelements, for instance tiles (blue rectangles in \cref {fig:tiles_and_blocks}(a)). Each tile is coded separately and contains several blocks (red border). In each tile, there is a \emph{reference block} (red block with yellow border) that is coded independently of the other blocks. Remaining blocks are encoded/decoded in a fixed order depicted by the black arrows. Then, upon request of a viewport, whose projection on the 2D image is depicted in green, all tiles containing at least some of the requested green area are sent and decoded. Note that the correlation between the tiles is not exploited and also potentially many unrequested blocks are sent. Therefore, these schemes cannot achieve the oracle compression rate of $\mathcal{R}^*$.

%
%
%

By contrast, in our scheme, the fixed decoding order constraint is relaxed. A block (rectangle with a red border in  \cref {fig:tiles_and_blocks}(b)) can be decoded using any combination of neighboring blocks that are available. 
Then, for each requested viewport, only the displayed blocks are decoded, and the correlation between the blocks is \emph{optimally} taken into account. 
Therefore,  our scheme can reach the  oracle compression rate $\mathcal{R}^*$, while maintaining the constraint that the request is not known in advance. 



\begin{figure}[h]
\centering
\includegraphics[width=1\linewidth]{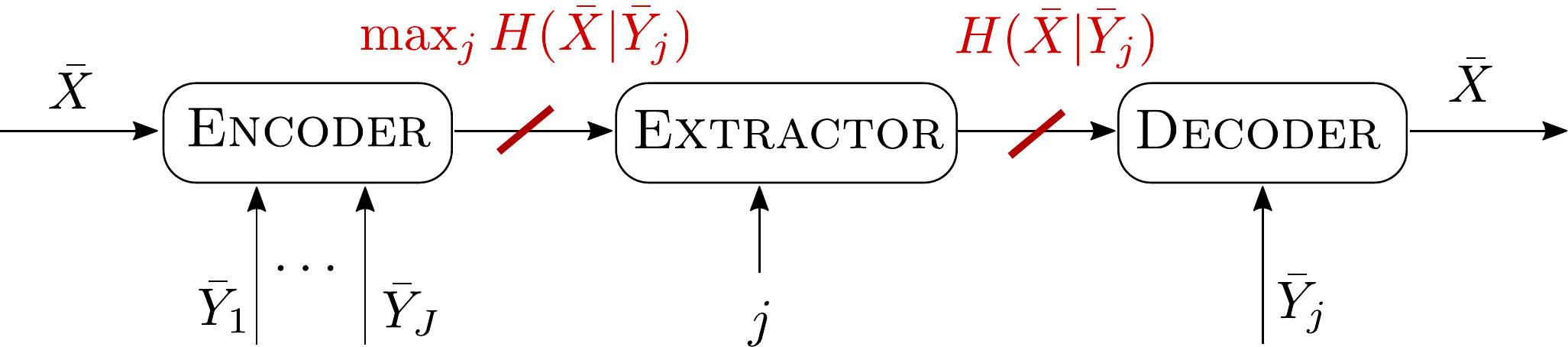}
\caption{
Compression rates achieved by incremental coding.
A source $\bar{X}$ is compressed knowing that a SI $\bar{Y}_j$ is available at the decoder. The encoder knows the set $\{\bar{Y}_i, i=1, \ldots, J\}$  but does not know which SI of the set is available at the decoder. The source can be compressed at rate $\max_i H(\bar{X}|\bar{Y}_i)$ corresponding to the conditional encoding with respect to the less correlated SI. From this description, a sub-description can be extracted at rate $H(\bar{X}|\bar{Y}_j)$, i.e. at the same rate as if the  SI was known by the encoder. 
}
\label{fig:Icoding}
\end{figure}

The aforementioned optimality can be explained from \cite{7351125}, which studies lossless coding of a source with several SI available at the encoder, see Fig.~\ref{fig:Icoding}. The source is encoded in a single bitstream and once it is known which SI is available at the decoder a part of the compressed bitstream is extracted.
In the 360$^\circ$ image context, $\bar{X}$ models the image block to be compressed and 
 each $\bar{Y}_j$ corresponds to the prediction for $\bar{X}$ using a combination of the neighboring blocks indexed by $j$.   
In \cite{7351125, dupraz2019rate, maugey2020incremental}, it is shown that \emph{incremental} coding \emph{optimally} exploits the dependence between the sources because the extraction can be made at rate $H(\bar{X}|\bar{Y}_j)$,\emph{ i.e.}, at the same rate as if the encoder knows in advance the index of the SI. Moreover, to achieve this optimality, there is no need to perform \emph{exhaustive storage} to store a bitstream for each possible SI, \emph{i.e.}, with a rate $\sum_j H(\bar{X}|\bar{Y}_j)$. Indeed, storing a bitstream for the less correlated SI at rate $\max_j H(\bar{X}|\bar{Y}_j)$ is sufficient. In particular, $\max_j H(\bar{X}|\bar{Y}_j)\ll \sum_j H(\bar{X}|\bar{Y}_j)$, especially when the number of possible SI is large. 

The proposed image interactive coding scheme works as follows. At the encoder, for each block (rectangle with red border in \cref{fig:tiles_and_blocks}(b)), a set of predictions is computed, one per set of neighboring blocks, that might be available at the decoder. Then, each block is encoded with the incremental encoder, conditionally to the set of SI. Then, the reference blocks of the tile-based approach are replaced by the so-called \emph{access blocks} (red shaded), for which, the set of predictions is complemented by a $\bar{Y}_{\emptyset}$, being an empty prediction. Therefore, this block can be decoded independently (as the first decoded block) or as a function of a the neighboring blocks if they are already decoded. In other words, these access blocks are stored with a rate $H(\bar{X})$ but can be transmitted at the proper rate, \emph{i.e.,} $H(\bar{X})$ if the block is the first block to be decoded and $ H(\bar{X}|\bar{Y}_j)$ if not.

At the decoder side, given a user request (green area in \cref{fig:tiles_and_blocks}(b)), an access block is decoded.  Then, from this block, a prediction for a neighboring block is calculated. This prediction is corrected thanks to the extracted bitstream that has been generated by the incremental source encoder, and this reconstructs the neighboring block, see Fig.~\ref{fig:Icoding}. 
Decoding of the following blocks is done in the same way until all blocks in the requested region are reconstructed. The decoding order is depicted as a black arrow in \cref{fig:tiles_and_blocks}(b).

%

%
%
%
%
%
%

%

\section{Detailed description of the proposed interactive coding scheme} \label{sec:details}

We now detail our interactive coding scheme. We first present the overall architecture, as well as the issues raised by the construction of an interactive coding scheme in Section~\ref{subsec:architecture}. These issues are further elaborated in the following sections.
For instance, we formulate the access block placement problem and propose a greedy solution in \cref{subsec:checkpoints_position}. We also explain the construction of the set of predictions in \cref{subsec:set_of_SIs}. We then detail the practical implementation of the incremental entropy coder, and its integration within an interactive image coder in  \cref{subsec:incrementalCoder}. Finally, we formulate the decoding order problem and propose a greedy solution in \cref{subsec:scan}. The way our proposed scheme handles navigation within a still scene is also discussed in \cref{subsec:scan}.

\subsection{Proposed architecture}
\label{subsec:architecture}

\begin{figure*}[h]
\centering
\includegraphics[width=0.65\linewidth]{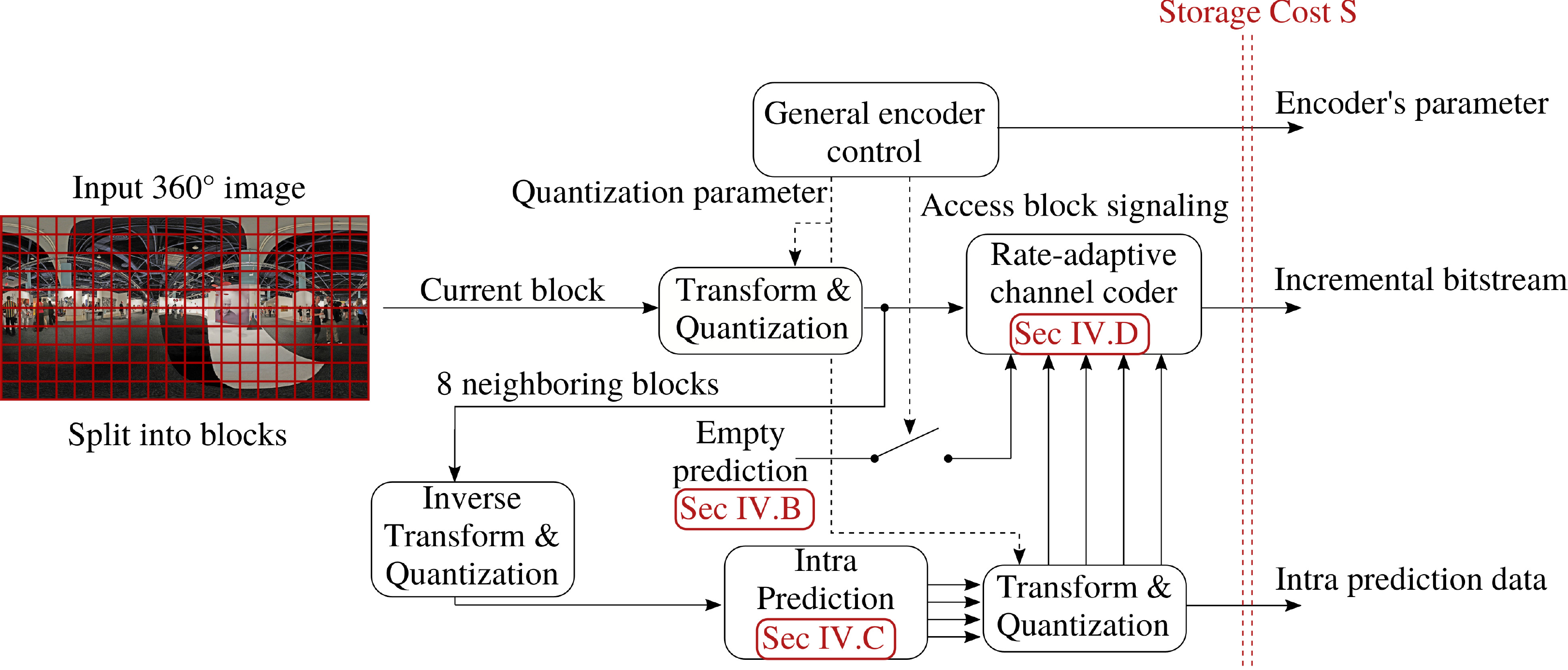}
\centerline{(a) Encoder}
\includegraphics[width=0.8\linewidth]{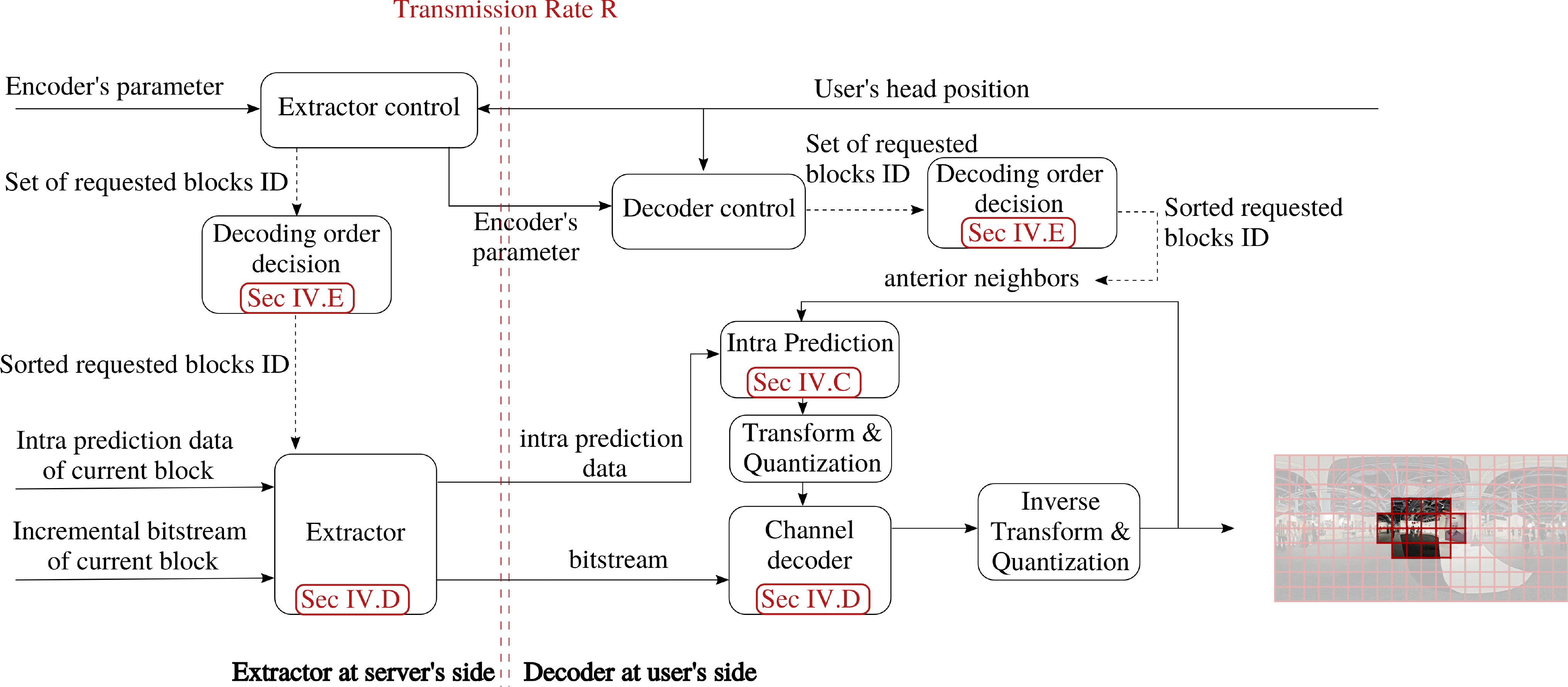}
\centerline{(a) Extractor and decoder}
\caption{Proposed architecture.}
\label{fig:proposedArchi}
\end{figure*}

The proposed architecture is summarized in Fig.~\ref{fig:proposedArchi}.
The encoder first splits the input image in several blocks $X_j$. 
For each block, several predictions are computed. There is one prediction per possible set of neighboring blocks available at the decoder (see \cref{subsec:set_of_SIs}). For some of these blocks, the so-called {access blocks} ($\{X_k\}_{k \in \mathcal{A}}$), \textit{i.e.} an empty prediction, is added to the prediction set, so that they can be decoded independently to start the decoding process. 
These blocks need to be carefully selected to ensure that any requested viewport can be decoded, but do not impact too much the storage  cost and transmission rate.
This trade-off is studied in \cref{subsec:checkpoints_position}. 
Then, each block $X_j$ is encoded with an incremental coder as a function of different predictions (see \cref{subsec:incrementalCoder}). This incremental coder, together with the quantizer and the transform must be carefully integrated in the image scheme to avoid the propagation of quantization errors \cite[Sec. 11.3]{sayood17b}. This is presented in \cref{subsec:incrementalCoder}.

At the decoder, given a requested viewport, one access block is first decoded. 
We optimize the block decoding order,  to reach a better storage and  transmission efficiency (see \cref{subsec:scan} for details). 
Then, each block is consecutively decoded by first generating a prediction based on the already decoded blocks, then extract the necessary bitstream, and run the incremental decoder (see \cref{subsec:incrementalCoder}).

We now turn to the optimization of the two critical features identified above: the access block set $\Ac$ and the decoding order, denoted $\tau$. 
Both features influence the storage cost $S(\Ac,\tau)$ and the transmission rate $R(\Ac,\tau)$, and must satisfy the constraint that any requested viewport must be decoded. To avoid sending unrequested blocks, and therefore minimize the transmission rate of the first request, we turn this constraint into the more stringent one:  
\begin{constraint}
\label{co:1ABperRequest}
There is at least one access block per request. More precisely,  with any rotation angle $\theta$ of the HMD device, there exists at least one access block belonging to the mapped visible region of the sphere $(m \circ \gamma)(\theta)$, \textit{i.e.,}
\begin{align}
\Cc(\Ac) = \left\{ \forall \theta, \ \exists \ k \in \mathcal{A} \  
\mbox{s.t.} \ X_k \in (m \circ \gamma)(\theta) \right\}.
\label{eq:1ABperRequest}
\end{align}
\end{constraint}
The new Constraint \ref{co:1ABperRequest} is more stringent than imposing that any request can be decoded, as it could increase the number of access blocks and therefore the storage. However, the proportion of access block among all blocks remains low (as shown in \cref{fig:checkpoints_location}). 
Therefore, storage is not significantly increased. 

As a consequence the access block/decoding order optimization problem can be formulated as 
\begin{align}
\min_{\Ac,\tau \mbox{ s.t. } \Cc(\Ac) } R(\Ac,\tau)+ \lambda S(\Ac,\tau),
\label{eq:pb_general}
\end{align}
where the constraint $\Cc(\Ac)$, corresponding to Constraint~\ref{co:1ABperRequest} is defined in \eqref{eq:1ABperRequest}. Furthermore, the storage does not depend on the
decoding order. Moreover, we have tested that, given a decoding order, the position of the access block has little influence on the transmission, see \cref{subsubsec:ABandR}. Therefore, the optimization problem \eqref{eq:pb_general} can be simplified into  two equivalent subproblems:

\begin{subequations}\label{eq:pb}
    \begin{empheq}[left={\displaystyle \min_{\Ac,\tau \mbox{ s.t. } \Cc(\Ac)} R(\tau)+ \lambda S(\Ac) \Leftrightarrow\empheqlbrace\,}]{align}
      & \min_{\Ac \mbox{ s.t. } \Cc(\Ac) }  S(\Ac) \\
      & \min_{\tau } R(\tau)
    \end{empheq}
\end{subequations}


\subsection{The access block placement problem} \label{subsec:checkpoints_position}

\begin{figure}[h!]
\centering
\subfloat[][]{\includegraphics[width=0.9\linewidth]{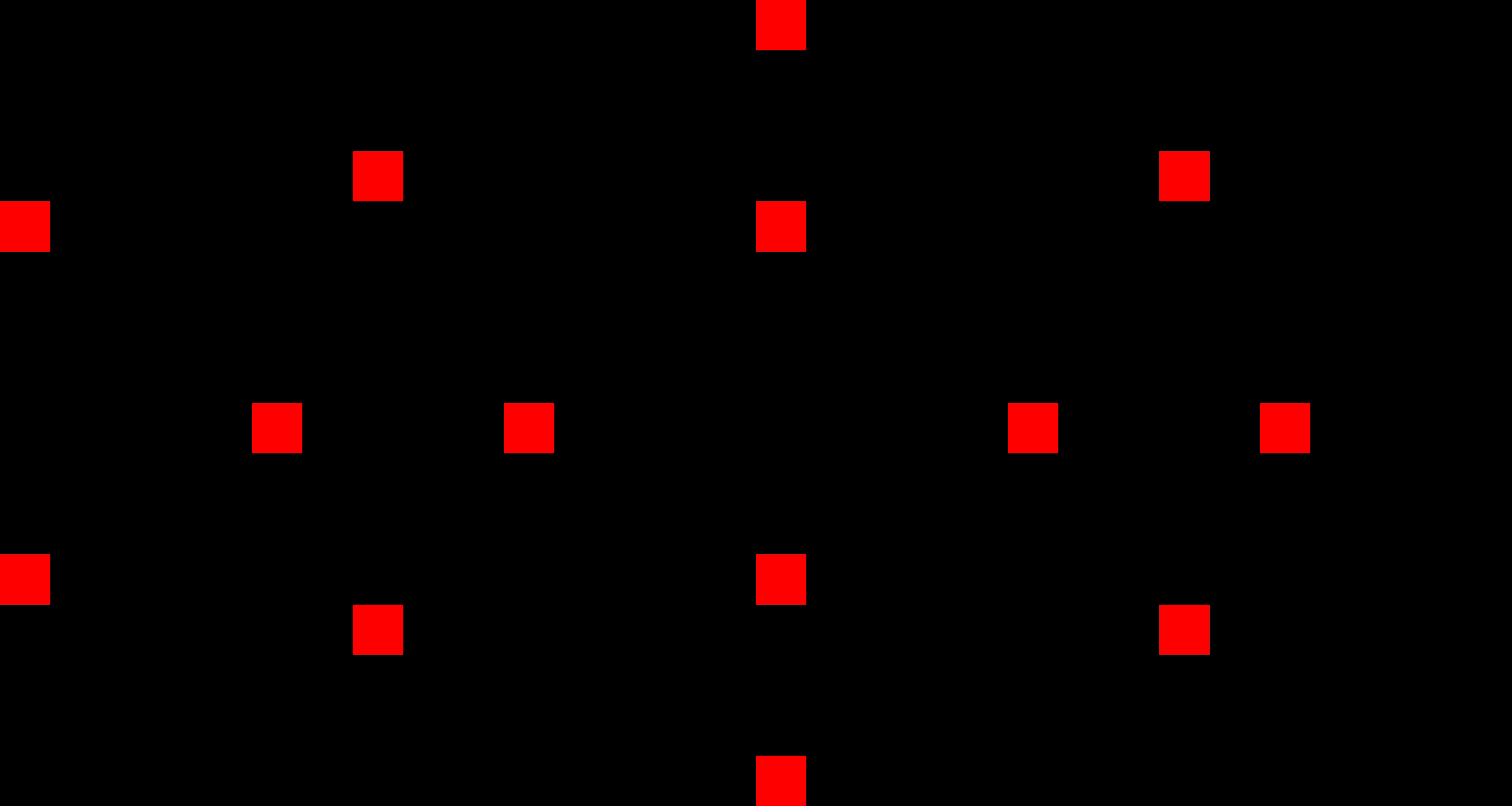} \label{subfig:fixed-based_access_block}}
\hfill
\subfloat[][]{\includegraphics[width=0.9\linewidth]{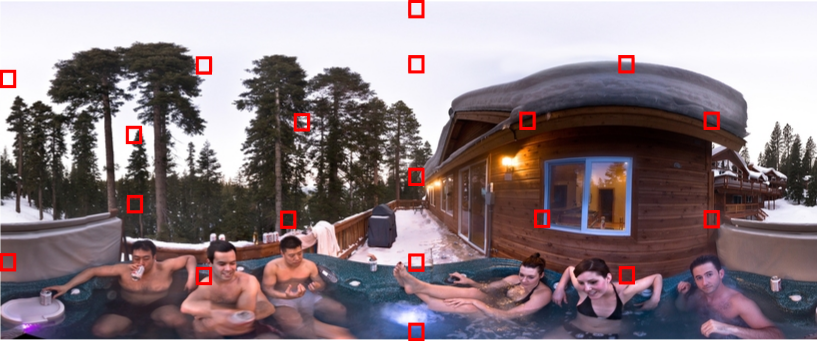}\label{subfig:content-based_access_block}}
\caption{Access block placement. (a) Fixed-based approach in which access blocks (in red) are selected without taking into account the content. (b) Content-based approach that takes into account the content to minimize the storage.}
\label{fig:checkpoints_location}
\end{figure}

To solve the access block placement problem (\ref{eq:pb}a), we propose two strategies. The first strategy is content independent and assumes that all blocks have the same storage cost. This algorithm is called \textbf{Fixed-based} approach in the following.
In other words, it approximates  (\ref{eq:pb}a) by 
\begin{align}\label{eq:pbAB1}
\min_{\Ac \mbox{ s.t. } \Cc(\Ac) } |\Ac|.
\end{align} 
The advantage of this strategy is that the solution only  depends on the spatial resolution and on the field of view of the HMD. As a consequence, there is no need to transmit the positions of the access blocks to the decoder. The new problem \eqref{eq:pbAB1}  is still complex since the constraint $\Cc(\Ac)$ is  a non-trivial function of the access block set. To solve it,  we propose a greedy algorithm.
Starting from the north pole, we change the angle $\theta$ to sweep the sphere to the south pole, and at each step, we check if an access block exists in $(m \circ \gamma)(\theta)$. If not, the block which contains the center of the area is added to the $\mathcal{A}$ set. The details of the algorithm can be found in Alg.~\ref{alg:access_point_blocks}. 
The obtained access block locations in the equirectangular image are shown in \cref{subfig:fixed-based_access_block}.

The second strategy instead takes into account the visual content and is called \textbf{Content-based}. In this strategy, first, the number of access blocks needs to be sent, which requires $\log_2(N)$ bits, if fixed-length encoding is used, and if $N$ represents the number of blocks in the image. Second, for each access block, the position of the block must be transmitted as well (again $\log_2(N)$ bits for each access block). If we denote the cost to encode the block of index $k$, independently of any other block by $R_{k|0}$, (\ref{eq:pb}a)  can now be rewritten as
\begin{align}\label{eq:pbAB2}
\min_{\Ac \mbox{ s.t. } \Cc(\Ac) } \log_2(N) + |\Ac| \log_2(N) + \sum_{k \in \Ac}R_{k|0} .
\end{align} 
Again, to solve this problem, we propose a greedy algorithm. It is similar to the previous one. The only difference is that, if no access block has been found in the current area, then the block with minimum rate is chosen. If many such blocks are found, then the one, which is closer to the center is chosen. The obtained access blocks positions are shown in \cref{subfig:content-based_access_block}. The impact of both approaches on the storage cost are evaluated in Sec.~\ref{sec:ablation_AB}.


\begin{myalgorithm}
\caption{Algorithm for placing access blocks}\label{alg:access_point_blocks}
\begin{algorithmic}[1]
\renewcommand{\algorithmicrequire}{\textbf{Input:}}
\renewcommand{\algorithmicensure}{\textbf{Output:}}
\REQUIRE ~~\\
$\theta = (\theta_{1}, \theta_{2})$ \\
$\theta_{1}$: longitude, $\theta_{1} \in [-\pi, \pi)$ \\
$\theta_{2}$: latitude, $\theta_{2} \in [-\pi/2, \pi/2]$ \\
$\Delta\theta_{1}, \Delta\theta_{2}$: predefined steps \\
$center(\theta)$: returns the block index in $m(\Gamma)$ which contains the center of $\gamma(\theta)$
\ENSURE  $\mathcal{A}$
\\ \textit{Initialization}: $\mathcal{A} = \emptyset$
\FOR{$\theta_{2}=-\pi/2$ \TO $\theta_{2} \leq \pi/2$ }
\FOR{$\theta_{1}=-\pi$ \TO $\theta_{1} < \pi$ }
\STATE $\theta = (\theta_{1}, \theta_{2})$
\IF {$\nexists \ k \in \mathcal{A} \ \mbox{s.t.} \ X_k \in (m \circ \gamma)(\theta)$ }
\STATE $\mathcal{A} \leftarrow \mathcal{A} \cup \{center(\theta)\}$
\ENDIF
\STATE $\theta_{1} \leftarrow \theta_{1} + \Delta\theta_{1}$
\ENDFOR
\STATE $\theta_{2} \leftarrow \theta_{2} + \Delta\theta_{2}$
\ENDFOR
\RETURN $\mathcal{A}$
\end{algorithmic}
\end{myalgorithm}

\subsection{Set of possible side information and prediction functions} \label{subsec:set_of_SIs}


The estimation of a given block by one or several of its neighbors is performed by adapting the intra-prediction module of conventional video coding standards \cite{wien2014high}. More precisely, the boundary row/column pixels from the adjacent blocks are combined to form a directional or planar prediction.
By construction, the prediction depends on which neighboring blocks are already decoded and available to serve as reference. In conventional video coding, the blocks are encoded and decoded with the raster scanning order, which means that the reference blocks are the left or top side. In our case, the order depends on the requested viewport and therefore, the decoder should be able to address any possible order. First, this decoding order influences which blocks are used as reference (also called context or causal information). Second, given these references, new prediction functions have to be defined.

\begin{figure}[h!]
\centering
\subfloat[][Type 1 SI]{
\begin{minipage}[b]{0.45\linewidth}
\def\svgwidth{1\linewidth} 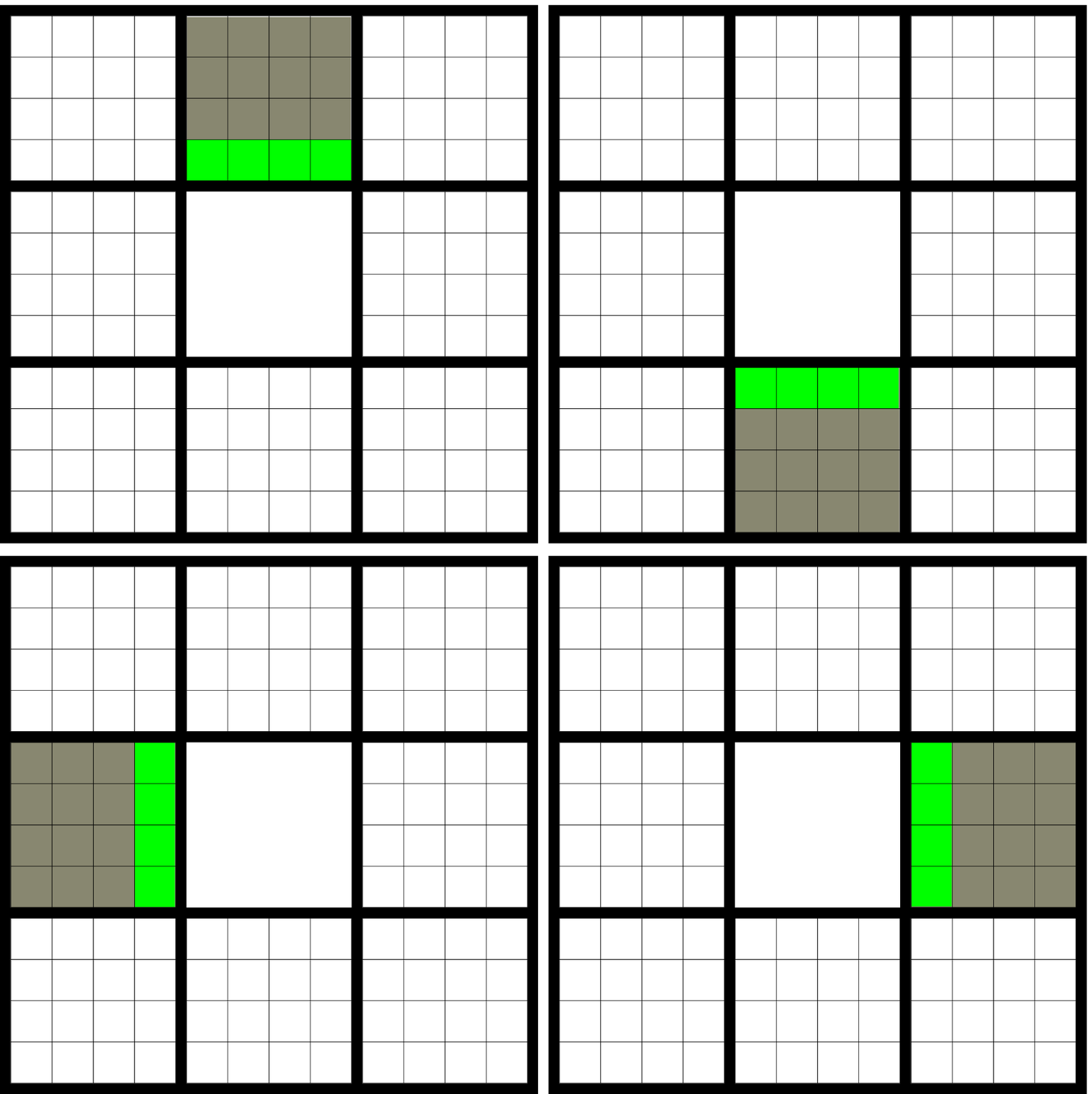\label{fig:SI_type1}
\end{minipage}
}%
\subfloat[][Type 2 SI]{
\begin{minipage}[b]{0.45\linewidth}
\def\svgwidth{1\linewidth} 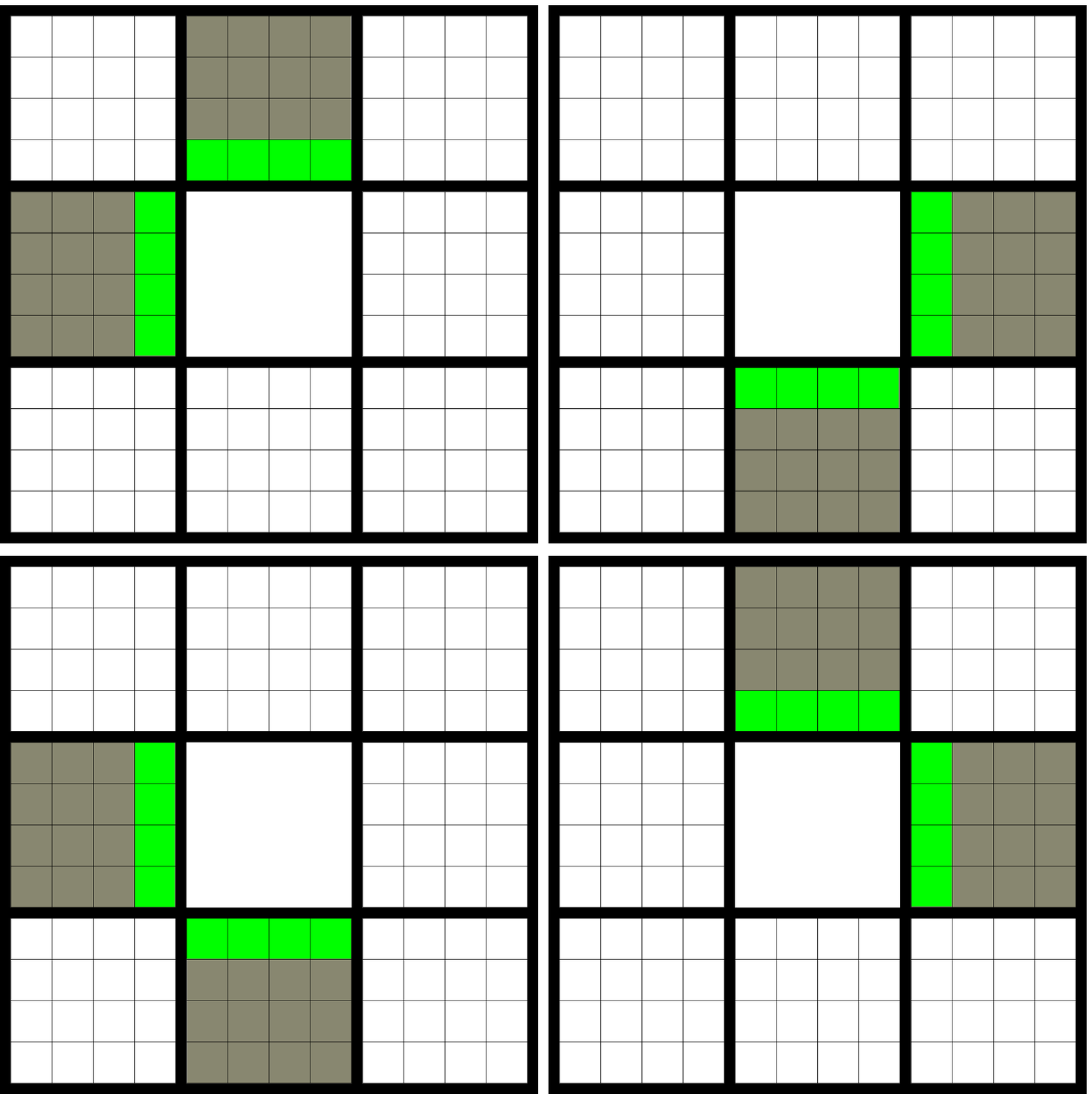\label{fig:SI_type2}
\end{minipage}
}%
\hfill
\subfloat[][Type 3 SI]{
\begin{minipage}[b]{0.45\linewidth}
\def\svgwidth{1\linewidth} 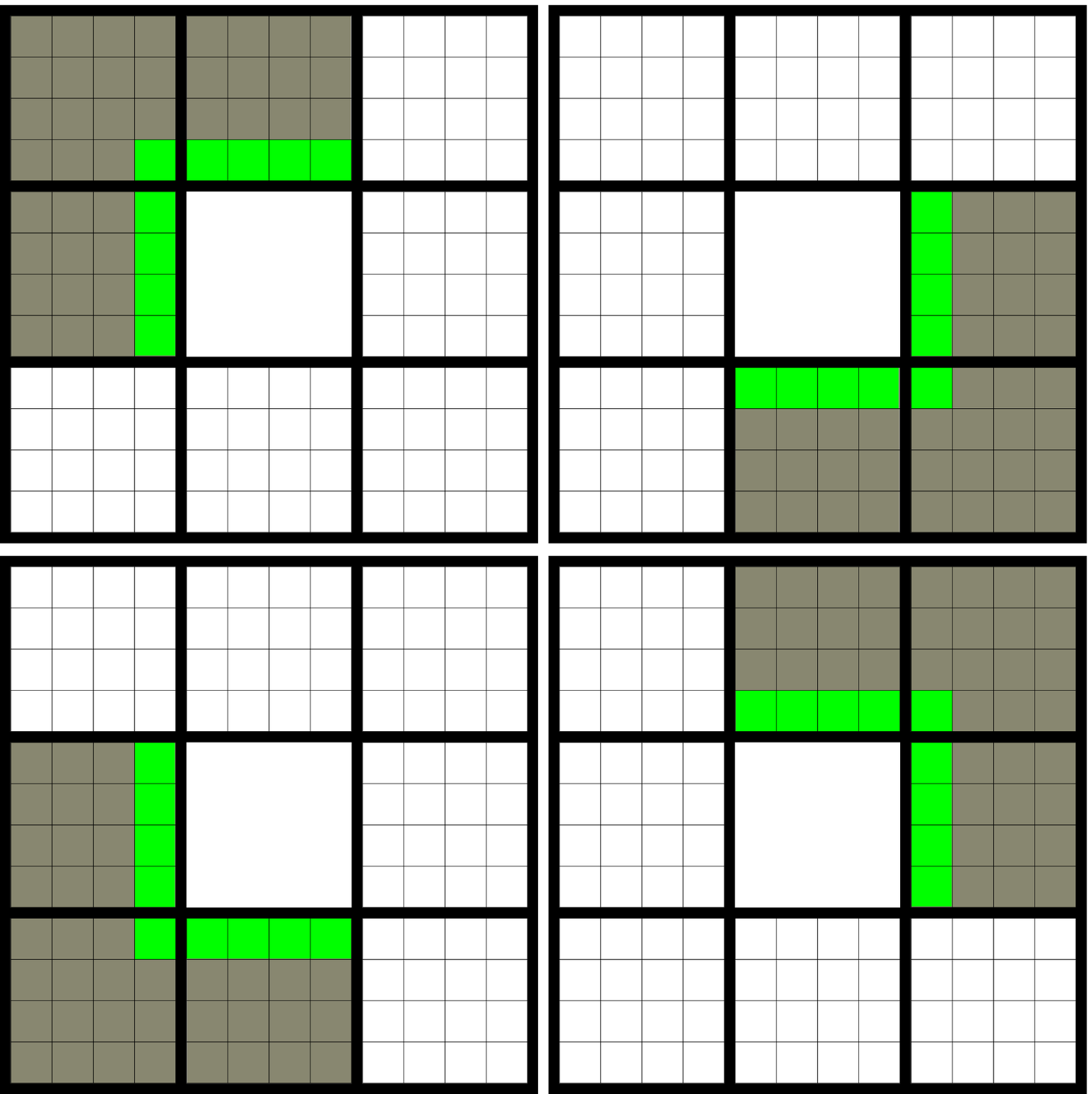\label{fig:SI_type3}
\end{minipage}
}%
\subfloat[][Intra prediction example]{
\begin{minipage}[b]{0.45\linewidth}
\centering
\includegraphics[width=1.1\linewidth]{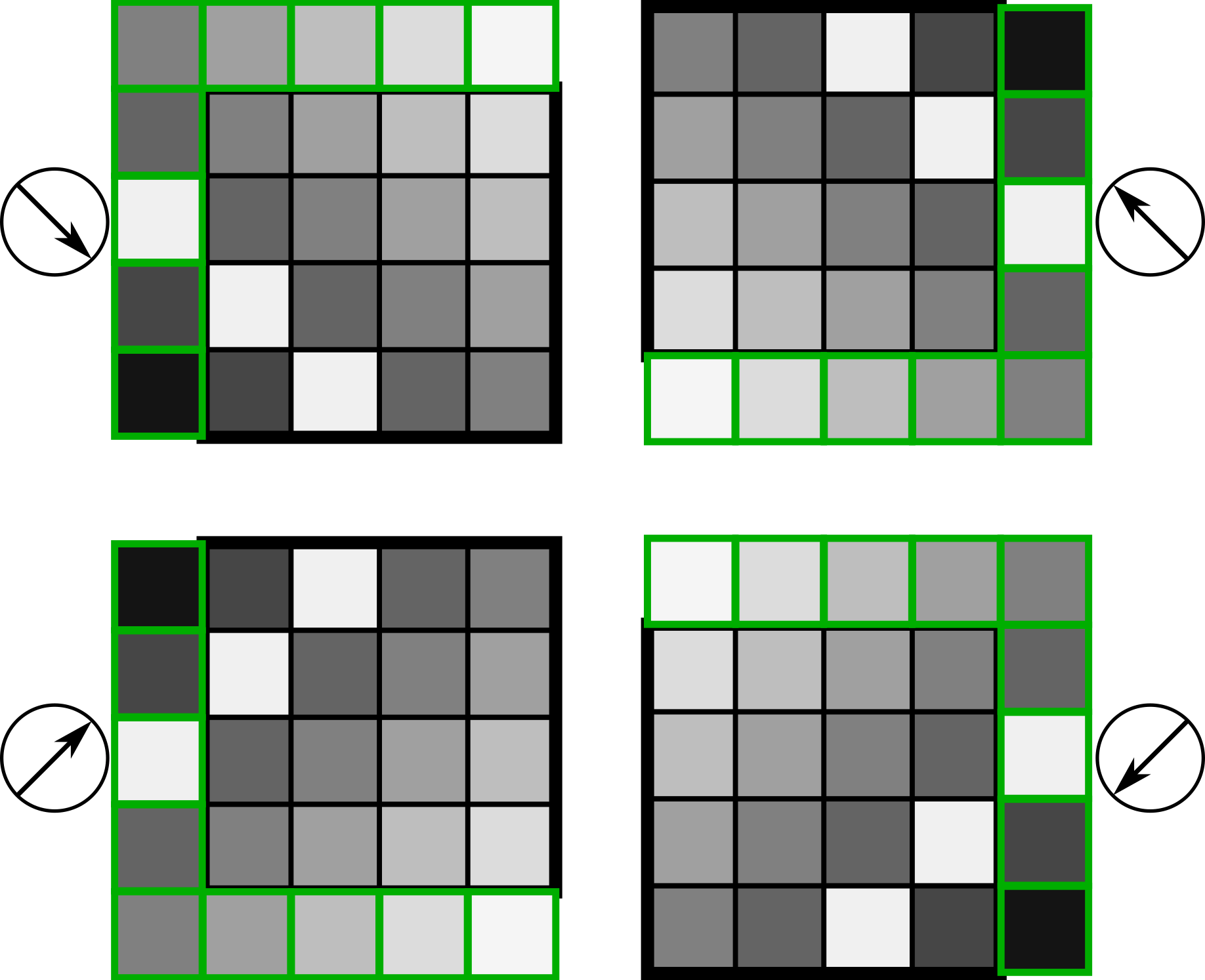}
\label{fig:SI_type3_intra}
\end{minipage}
}%
\caption{Set of possible SI per block. (a-c) Reference pixels (in green) used to predict the block $X$ for each prediction type. The prediction serves as SI for the entropy coder. The available neighboring blocks are depicted in dark gray. (d) Intra-prediction examples for type 3.}
\label{fig:side_info_types}
\end{figure}

%
The blocks that can be used as reference depend on the decoding order. We categorize them according to the number of neighboring blocks available at the decoder. This leads to 3 types, as illustrated in \cref{fig:side_info_types}:
\begin{itemize}
\item \textit{Type 1}: Only one of the adjacent blocks has been decoded.
\item \textit{Type 2}: A vertical and a horizontal border at the boundaries are available to perform the prediction. This means that a block at the top or bottom and a block at the right or left side of the current block have been decoded already.
\item \textit{Type 3}: This type is similar to \textit{type 2} except that, in addition to the vertical and horizontal border, the corner block between them has also been decoded.
\end{itemize}
For each type, there exist 4 states, which leads to 12 possible contexts in total. Each context gives rise to a possible prediction. 
Prediction functions are deduced from the prediction modes of conventional video coding standards by a rotation of multiples of 90$^\circ$. Examples of type 3 SI predictions are shown in \cref{fig:SI_type3_intra} with their corresponding prediction mode.

In general, if there is no limitation for the navigation, all of these $12$ predictions are generated and used to encode the block $X$. For the blocks which are on the top and bottom rows of the equirectangular image, since navigation can not come from the top or bottom side respectively, the non-available side information is removed from the possible set. By contrast, for the blocks which are on the left and right boundaries of the equirectangular image, since when they are wrapped to the surface of the sphere they are adjacent to each other, they can be used as a reference to predict the opposite side. For example, for a block which is located at the rightmost part of the equirectangular image, the leftmost block at the same row can be used as a reference for its prediction and vice versa.

The predictions for a block $X$ are denoted by $Y_1, Y_2, \ldots, Y_L$ in the following, and each prediction corresponds to one possible SI. In order to be consistent at both encoder and decoder, and avoid propagation of quantization errors, the predictions are generated with the lossy (quantized) version of the neighboring blocks.

\subsection{Practical implementation of incremental source code}\label{subsec:incrementalCoder}

A practical solution to implement incremental codes can be achieved by rate-adaptive channel codes such as rate-adaptive 
Low-Density Parity-Check (RA-LDPC). Channel codes have been used in other interactive coding schemes like stream-switching \cite{5167460, cheung2008compression}, where distributed source coding (DSC) is used to merge multiple SI frames. More precisely, and similar to our context, a set of SI is computed at the encoder and only one of them is available at the decoder. However, the current block is encoded
by exploiting the worst-case correlation between a set of potential SI frames and the target frame. Here, thanks to the theoretical work obtained in \cite{7351125, dupraz2019rate, maugey2020incremental} for the incremental code, we propose to store the data w.r.t. the worst-correlated prediction (or equivalently SI), but transmit optimally by extracting the amount of data required to decode based on what is available at the decoder.

\begin{figure*}[h!]
\centering
\fontsize{7pt}{7pt}\selectfont
\def\svgwidth{0.99\linewidth}
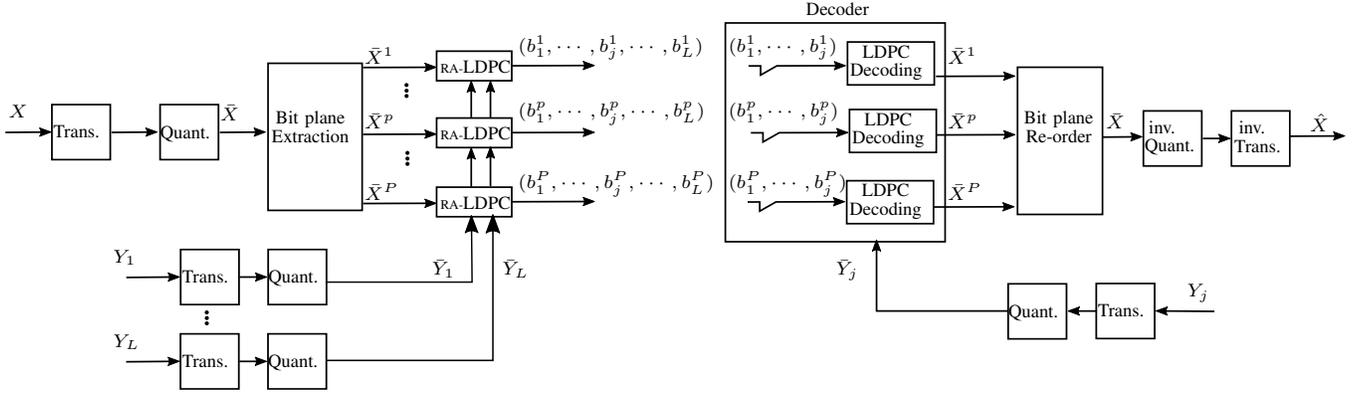 
\caption{Lossy coding scheme of an image block $X$ as a function of its predictions $Y_1,\ldots, Y_L$. }
\label{fig:detailed_coding_scheme}
\end{figure*}

The proposed incremental coding scheme for encoding and decoding a block $X$ is depicted in \cref{fig:detailed_coding_scheme}.
The input block $X$ is first transformed and quantized resulting to a signal $\bar{X}$ that is split into several bitplanes $\bar{X}^1, \ldots, \bar{X}^P$. We now assume that the predictions $Y_1, \ldots, Y_L$ are sorted from the best to the worst, \emph{i.e.,} $Y_1$ is more  correlated with $X$ than $Y_L$. These predictions are also transformed and quantized resulting to $\bar{Y}_1, \ldots, \bar{Y}_L$. Note that doing so, encoder and decoder are consistent, which avoids the propagation of quantization errors \cite[Sec. 11.3]{sayood17b}.  

 Each bitplane $\bar{X}^p$ is encoded thanks to a binary RA-LDPC encoder, with $\bar{Y}_1, \ldots, \bar{Y}_L$ as SI sequences. It is worth noting that since the correlations between the source and its SI predictions are known in advance at the encoder, the symbol-based ($Q$-ary) and binary-based  RA-LDPC (with bitplane extraction) perform equally \cite{4379081}, but binary-based  RA-LDPC is less complex than its symbol-based version. This generates $L$ bitstreams $(b_1^p, \ldots, b_L^p )$ that are stored on the server.
For the decoding of $X$, if the prediction $Y_j$ is available, the bitstreams that need to be transmitted are $(b_1^p, \ldots, b_j^p )$. It is noticeable that a bitstream is not only dedicated to one particular prediction, but can serve the decoding of several predictions. This is the reason why the storage cost remains limited. Indeed, the stored bitstream does not depend on the number of possible SI $L$, but only on the source $X$ correlation with the worst SI.

For the access blocks, an additional variable $Y_{\emptyset}$ (a source of zero entropy) is considered as possible SI. Then, the encoding process is identical, which means that on top of $(b_1^p, \ldots, b_L^p )$, a bitstream $b_{\emptyset}^p$ is added  to complement the decoding in the case where the block has to be decoded independently. Note that the  bitstream $b_{\emptyset}^p$ serves as a complement of the other bitstreams in case of independent decoding, which means that $\sum_{j=1}^L |b_j^p| + |b_{\emptyset}^p| = H(\bar{X}^p)$.




\subsection{Decoding  order and navigation}\label{subsec:scan}

Blocks are decoded one after the other until all blocks in the requested area are processed. 
The choice of the decoding order is important since it impacts the quality of the predictions that are made, and therefore the transmission rate. More formally, let $\mathcal{J}(\theta)$ denote the set of displayed blocks by an HDM pointing in the direction $\theta$, \emph{i.e.,} $\mathcal{J}(\theta) = \{ j \ | \ X_j \in (m \circ \gamma)(\theta) \}$.  We seek for a decoding order, \textit{i.e.}, a permutation $\tau$ of this set such that the transmission rate is minimized as mentioned in (\ref{eq:pb}b). $\tau(j)$ stands for the position of $j$ in the new arrangement.  
The difficulty of this problem is that the rate to encode a block from another block depends on the number of neighbors and therefore depends on the decoding order. Therefore, the cost of the transition between two blocks is not fixed and the problem can not be formulated as an optimization over a graph. So, to solve this difficult problem, we propose a greedy approach, called \textbf{GreedyRate}. At each iteration, given the set of already decoded blocks, the one hop neighborhood around all decoded blocks is computed. This neighborhood corresponds to the blocks that can be decoded next. Among this set, the block with minimum rate is chosen.

However, \textbf{GreedyRate} needs to know the encoding rates of each block for each possible SI. Therefore, this algorithm can only be implemented at the encoder and the decoding order needs to be sent to the decoder. To avoid sending the decoding order and further save bandwidth, we replace the rate in  (\ref{eq:pb}b) by a heuristic, which can be computed at both encoder and decoder.
Indeed, the quality of the prediction, which impacts the transmission rate,  increases with the number of neighboring  blocks that serves as reference (Type 1, 2 or 3 in \cref{fig:side_info_types}). 
Therefore, we propose to find a decoding order that maximizes the overall number of blocks used for predictions, \textit{i.e.},
\begin{equation}
\max_{\tau} \ \sum_{j \in \mathcal{J}} |\{ j'\ | \ \tau(j') < \tau(j), X_j \sim X_{j'} \}|,
\end{equation}
where $\sim$ means that two blocks are neighbors.

Moreover, the experiments conducted in Sec.~\ref{sec:ablation_hor_vert}, show that the correlation between horizontal blocks is higher than between vertical blocks, \textit{i.e.}, the prediction that is generated by the left or right block generates less distortion than the prediction generated from the top or bottom block. Therefore, we augment the criterion to favor horizontal predictions
\begin{align}
\max_{\tau} \ \sum_{j \in \mathcal{J}} & |\{ j'\ | \ \tau(j') < \tau(j), X_j \sim X_{j'} \}| + \mathbbold{1}_{\mathcal{E}}(j) 
\label{eq:opt_scan}\\
\mathcal{E} =& \{ j | \exists j', X{j'} \sim X_j,  X_{j'} \in \mbox{ same line as } X_j \} \nonumber
\end{align}
where $\mathbbold{1}_\mathcal{E}(j)$ is the characteristic function of the set $\mathcal{E}$. Note that if the image is rotated, then the highest correlation may not be between horizontal blocks but rather vertical for instance. This direction of highest correlation can then easily be signalized with a single additional bit.
We propose a greedy solution to \eqref{eq:opt_scan} which is called \textbf{GreedyCount}.
At each iteration, given the set of already decoded blocks, the one hop neighborhood is computed. This neighborhood corresponds to the blocks that can be decoded next. Among this set, the blocks with maximum number of neighbors are identified. Among this subset, we select a block that has maximum number of horizontal transitions.

To further reduce the complexity of the \textbf{GreedyCount} algorithm, we propose the \textbf{SnakeLike} inspired by the one-scan connected component labeling technique \cite{4728561}, where we add a preference to horizontal transitions. More precisely, the input of the algorithm is the set of requested blocks and the output is the ordered sequence of these blocks. At each iteration, among the neighbors of the last selected block, a horizontal neighbor is chosen. If there is no horizontal neighbor (either because the horizontal neighbor is not requested or because it is already in the ordered list), then a vertical neighbor is chosen. If there is no neighbor, then previous process is done for the last selected block with a neighbor. 
The proposed solution
can start at any position of a block (not necessarily at a corner) and is adapted to any shape (not necessarily a rectangle). These latter adaptations are made possible through a recursive implementation of the scanning order. An example of the snake scan  is shown with black arrows in \cref{fig:tiles_and_blocks}(b).

Navigation within a 360$^\circ$ image is also handled by our interactive coder. Indeed, navigation consists in requesting a sequence of directions or equivalently a sequence of viewports. Therefore, for any newly requested viewport, the already decoded blocks of this viewport and its boundary are determined. One of these blocks is chosen as the starting point. A decoding order is computed for the remaining non-decoded blocks. In conclusion, our scheme handles efficiently navigation by sending only the unknown blocks and even avoiding to send a block as an access block.  

To summarize, three algorithms have been proposed to approach the difficult decoding order optimization problem  (\ref{eq:pb}b). \textbf{GreedyRate} is based on the rates and is thus content based. The two others \textbf{GreedyCount} and \textbf{SnakeLike} algorithms use a heuristic and do not depend on the image.


%

\section{Experimental results} \label{sec:experiments}

\begin{figure*}[htb!]
    \centering
\subfloat[][Market]{\includegraphics[width=0.24\textwidth, height=0.12\textwidth]{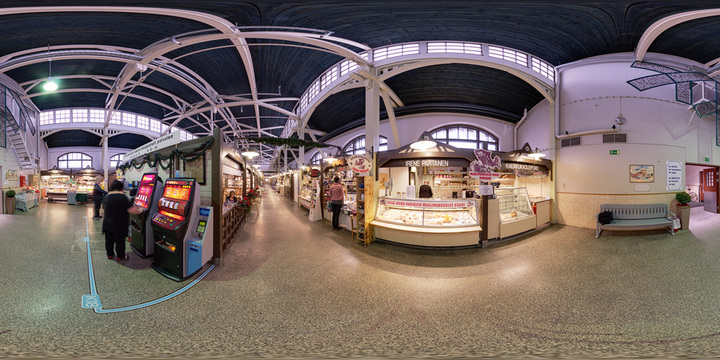}}
\hfill
\subfloat[][Street]{\includegraphics[width=0.24\textwidth, height=0.12\textwidth]{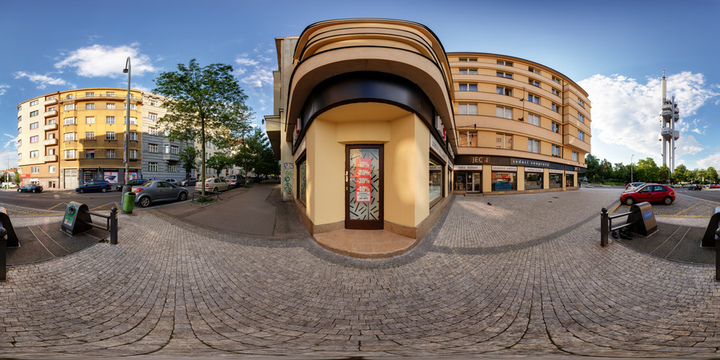}}
\hfill
\subfloat[][Mountain]{\includegraphics[width=0.24\textwidth, height=0.12\textwidth]{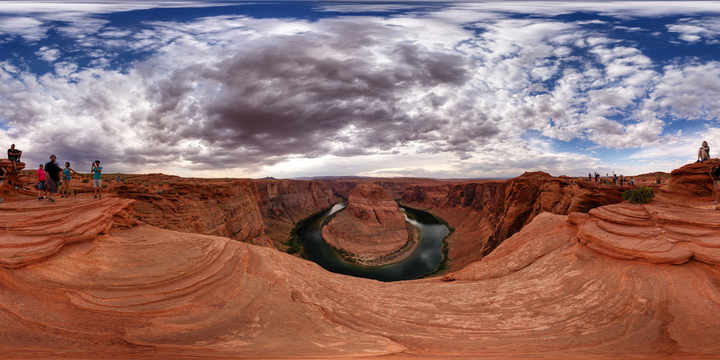}}
\hfill
\subfloat[][Church]{\includegraphics[width=0.24\textwidth, height=0.12\textwidth]{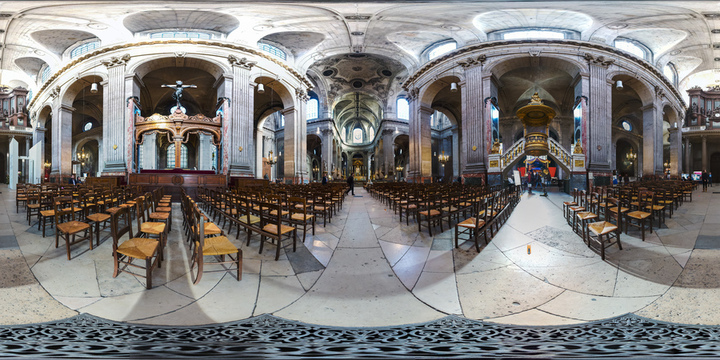}}
\hfill
\subfloat[][Seashore]{\includegraphics[width=0.24\textwidth, height=0.12\textwidth]{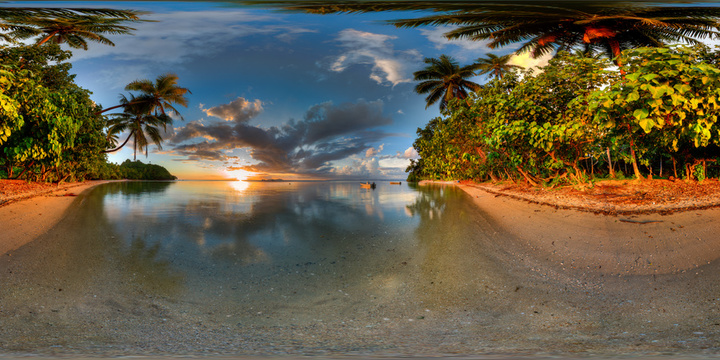}}
\hfill
\subfloat[][Park]{\includegraphics[width=0.24\textwidth, height=0.12\textwidth]{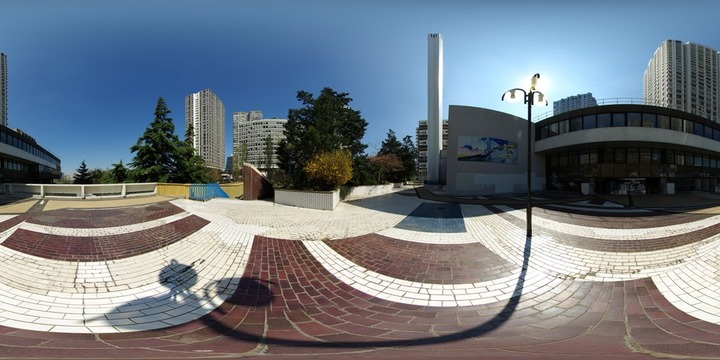}}
\hfill
\subfloat[][Jacuzzi]{\includegraphics[width=0.24\textwidth, height=0.12\textwidth]{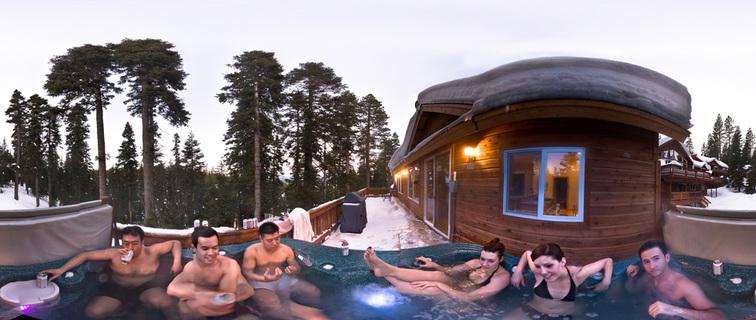}}
\hfill
\subfloat[][Cafe]{\includegraphics[width=0.24\textwidth, height=0.12\textwidth]{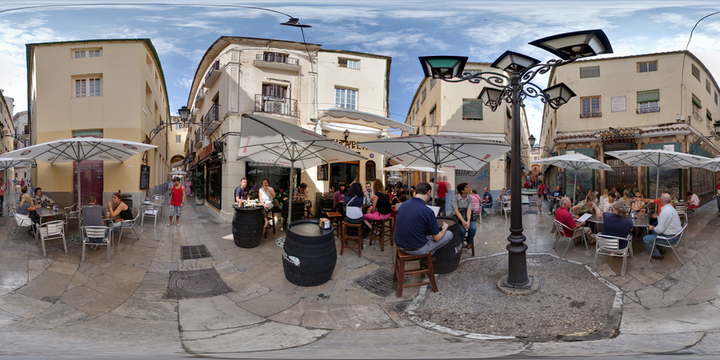}}
\caption{Images used in our experiment and their corresponding names.}
\label{fig:experiment_images}
\end{figure*}



For the experiments, we selected 8 large equirectangular images from the Nantes University dataset \cite{rai2018}. We have named images for the sake of clarity. These images and their corresponding names are shown in \cref{fig:experiment_images}. In this dataset, each 360-degree image has been viewed by several observers. The head movements have been recorded as trajectories on the unit sphere. For each sampled head position, the longitude and latitude values of the center of the viewport are presented. The collected head positions have been down-sampled over sequential windows of 200 msec. The head movement data can be also predicted using the salient areas \cite{ZHU201815}. For our experiment, the blocks are of size $32\times 32$. The rate-adaptive channel code used in our scheme is the RA-LDPC \cite{8662618}.



In the next three subsections, we present different experiments to validate the choices made in the design of our coder. After these ablation studies we evaluate our coder with different baselines.
For that, we consider two coding schemes as baselines to evaluate the proposed coder, namely tile-based scheme and exhaustive storage. 

Different tile structures are used by partitioning the equirectangular image into $7\times 7$, $7\times 5$, $5\times 5$, $5\times 4$, $4\times 4$, and $2\times 2$ non-overlapping tiles. We denote them by ``T. $m \times n$" where $m$ and $n$ are the numbers of vertical and horizontal tiles, respectively. Inspired by \cite{7892357,8626197} we also consider an irregular tiling structure as another baseline in which the top 25\% of the image is partitioned into one tile, the bottom 25\% to another, and the middle 50\% is divided into four equal-area tiles. We denote this baseline by ``T. opt structure". We also consider the case when the whole equirectangular image is sent in the first request (T. $1\times 1$). 

The exhaustive storage  (ES) approach does not really exist in practice, but it illustrates the intuitions of some baselines adopted in other contexts \cite{kuzyakov2016next,7532582,6738368,5557810,1218196}.
As in our coding scheme, the ES scheme considers all the predictions for each block, but stores a residue for each of them. It is thus well adapted for transmitting at the oracle transmission rate $\mathcal{R}^*$, but has a great storage expense.

The storage cost is computed for each image and each baseline. The transmission rate and distortion are computed per user and per request while users are navigating in the scene using the recorded head positions. As explained in \eqref{eq:distortion}, the distortion is computed at each viewport of a user's request \cite{7328056}.

\subsection{Ablation study: access block}

\subsubsection{Access block position does not impact transmission rate}
\label{subsubsec:ABandR}
In this experiment, we prove that the transmission rate is not dependent on the location of access blocks. In other words, if we change the position of access block inside the requested area to start decoding the requested blocks, it doesn't impact the transmission rate. For that, we take the first request of every user in all images, and we change the position of the access blocks inside the requested area. Note that in this scenario, we assume that every requested block can be an access block, so this is without optimizing the position of access blocks, which we investigate in the next ablation study (\textit{Access block positioning}). We choose $K=100$ different positions for each request, and then start the decoding process from the chosen access block. We then compute the ratio between the Maximum Absolute Deviation of the transmission rates and the average transmission rate for two different quantization parameters (QP=22, 42). More precisely, for each QP and each user, if we denote the set of transmission rates of $K$ experiments with $\Rc_{\Ac} = \{r_1, r_2, \ldots, r_K\}$, we compute 

\begin{equation}
\begin{gathered} 
\frac{\max(\{ |r_i - \bar{r}|: i=1, \ldots, K\})}{\bar{r}}, \\
\text{where } \bar{r} = \frac{1}{K} \sum_{i=1}^{K} r_i.
\label{eq:ratio_maximum_absolute_deviation}
\end{gathered}
\end{equation}

The results are shown in \cref{tab:transmission_rate_vs_access_block}. It can be seen that in all cases, the maximum absolute deviation of the transmission rate is less than $0.1\%$ with respect to the average. Therefore, we can conclude that the \emph{transmission rate is independent of the position of access block}.

\begin{table}[]
\centering

\caption{The impact of access block position on the transmission rate. The average, maximum, and minimum values of \eqref{eq:ratio_maximum_absolute_deviation} over all users is computed for each QP.}
\label{tab:transmission_rate_vs_access_block}
\begin{tabular}{l|l|l|l|}
\cline{2-4}
 & Average & Maximum & Minimum \\ \hline
\multicolumn{1}{|l|}{Market} & 3.7e-04 & 1.1e-03 & 1.5e-04 \\ \hline
\multicolumn{1}{|l|}{Street} & 4.6e-04 & 7.3e-04 & 1.6e-04 \\ \hline
\multicolumn{1}{|l|}{Mountain} & 2.1e-04 & 5.7e-04 & 1.1e-04 \\ \hline
\multicolumn{1}{|l|}{Church} & 3.4e-04 & 7.8e-04 & 1.9e-04 \\ \hline
\multicolumn{1}{|l|}{Seashore} & 1.3e-04 & 3.0e-04 & 5.7e-05 \\ \hline
\multicolumn{1}{|l|}{Park} & 3.9e-04 & 8.5e-04 & 2.0e-04 \\ \hline
\multicolumn{1}{|l|}{Jacuzzi} & 5.5e-04 & 9.8e-04 & 2.7e-04 \\ \hline
\multicolumn{1}{|l|}{Cafe} & 5.1e-04 & 8.7e-04 & 2.5e-04 \\ \hline
\end{tabular}%

\end{table}

\subsubsection{Access block positioning}\label{sec:ablation_AB}
We compare the storage cost of the two algorithms proposed for access block positioning (see \cref{subsec:checkpoints_position}). Results are shown in \cref{tab:CA_vs_F_AB} for two QP values. The Content-based approach depends on the content, and therefore, there is a need to signal the position of access blocks. As can be seen from the table, even though the content-based approach provides lower storage, when we add the signalization cost, the overall storage cost exceeds the Fixed-based approach. Therefore, the Fixed access block positioning is sufficient for our purpose, plus it has the advantage that it is fixed between images with the same resolution. For the rest of this paper, we consider the Fixed-based approach for our access block positioning.

\begin{table}[]
\centering

\caption{Storage cost of Content-based and Fixed-based access block positioning. The results are shown in Kilo Byte (KB) and for two different QPs.}
\label{tab:CA_vs_F_AB}
\begin{tabular}{l|c|c|c|c|}
\cline{2-5}
 & \multicolumn{2}{c|}{QP=22} & \multicolumn{2}{c|}{QP=42} \\ \cline{2-5} 
 & ContentBased & Fixed & ContentBased & Fixed \\ \hline
\multicolumn{1}{|l|}{Market} & 13875.44 & \textbf{13875.41} & 8842.01 & \textbf{8841.97} \\ \hline
\multicolumn{1}{|l|}{Street} & 13768.33 & \textbf{13768.29} & 8057.99 & \textbf{8057.96} \\ \hline
\multicolumn{1}{|l|}{Mountain} & 11833.89 & \textbf{11833.85} & 6838.97 & \textbf{6838.93} \\ \hline
\multicolumn{1}{|l|}{Church} & 13102.20 & \textbf{13102.17} & 7235.98 & \textbf{7235.94} \\ \hline
\multicolumn{1}{|l|}{Seashore} & 13070.45 & \textbf{13070.41} & 7330.30 & \textbf{7330.26} \\ \hline
\multicolumn{1}{|l|}{Park} & 8597.98 & \textbf{8597.95} & 6066.15 & \textbf{6066.12} \\ \hline
\multicolumn{1}{|l|}{Jacuzzi} & 7453.13 & \textbf{7453.09} & 5141.92 & \textbf{5141.90} \\ \hline
\multicolumn{1}{|l|}{Cafe} & 8243.73 & \textbf{8243.70} & 5323.09 & \textbf{5323.04} \\ \hline
\end{tabular}

\end{table}

\subsection{Ablation study: decoding order}
\subsubsection{Horizontal/vertical prediction}\label{sec:ablation_hor_vert}
To confirm the validity of \eqref{eq:opt_scan}, it is sufficient to compare the compression rate of a block when only one side of the block is available as a reference to do prediction, \textit{i.e.,} comparing the compression rate between the predictions coming from Type 1 SI of \cref{fig:side_info_types}. For every block of all images, we measure the average compression rate of the block when the prediction comes from the left or right side (horizontal preference) and compare them with the average compression rate when the prediction comes from the top or bottom side of the block (vertical configuration). \cref{tab:horizontal_vs_vertical_prediction} shows the percentage of blocks for which the average compression rate obtained from the right and left predictions is lower than the average compression rates obtained from the top and bottom predictions  when QP=22. Similar behavior observed for other QP values. It can be seen that for all images more than 50\% of the blocks have lower compression rates when the predictions come from the left or right sides. This clearly shows that for decoding order scanning, it is preferable to favor horizontal predictions than vertical one \textit{when the camera orientation is horizontal} as it is the case in our test experiments.

\begin{table}[]
\centering

\caption{Percentage of blocks when the average compression rate of the horizontal predictions (the predictions come from the left or right side) is better than the vertical predictions (the predictions come from the top or bottom sides), and vice-versa. The remaining percentage corresponds to the case when both are equal. The results are related to QP=22.}
\label{tab:horizontal_vs_vertical_prediction}
\begin{tabular}{l|c|c|}
\cline{2-3}
 & \begin{tabular}[c]{@{}c@{}}Horizontal is better\\ (in \%)\end{tabular} & \begin{tabular}[c]{@{}c@{}}Vertical is better\\ (in \%)\end{tabular} \\ \hline
\multicolumn{1}{|l|}{Market} & \textbf{58.6\%} & 36.0\% \\ \hline
\multicolumn{1}{|l|}{Street} & \textbf{56.3\%} & 36.0\% \\ \hline
\multicolumn{1}{|l|}{Mountain} & \textbf{56.8\%} & 34.6\% \\ \hline
\multicolumn{1}{|l|}{Church} & \textbf{52.9\%} & 41.3\% \\ \hline
\multicolumn{1}{|l|}{Seashore} & \textbf{55.9\%} & 37.2\% \\ \hline
\multicolumn{1}{|l|}{Park} & \textbf{61.2\%} & 32.7\% \\ \hline
\multicolumn{1}{|l|}{Jacuzzi} & \textbf{52.8\%} & 35.9\% \\ \hline
\multicolumn{1}{|l|}{Cafe} & \textbf{55.3\%} & 39.1\% \\ \hline
\end{tabular}

\end{table}

\subsubsection{Performance of different decoding order algorithms}\label{subsec:AblaDO}
We conduct an experiment to analyze the performance of the three different decoding orders presented in \cref{subsec:scan}. For that, we compute the transmission rate for the first request of all users. The transmission rates averaged over all users are presented in \cref{tab:different_decoding_orders}. In the GreedyRate algorithm, we need to signalize the decoding order. Therefore, although it tries to optimize the transmission rate, the addition of the decoding order signalization results in a drop in the performance of the transmission rate. It is interesting to say that this drop of performance is not only because of signalization but also because this greedy algorithm can not always manage to find the lowest transmission rate compared to the other two methods. In fact, \emph{even if we do not consider the signalization cost of greedy algorithm}, only 62\% of the cases this GreedyRate has a lower transmission rate than our SnakeLike approach, and for the remaining 38\% of cases, SnakeLike algorithm performs better. This clearly shows the complexity of this optimization problem. However, if we consider the signalization cost, the GreedyRate approach always performs worse. Regarding the comparison between SnakeLike and GreedyCount approaches, we can clearly see that at the cost of a more complex algorithm, the GreedyCount performs slightly better than the SnakeLike approach. However, this efficiency is negligible. Therefore, for the rest of this paper, we consider the SnakeLike algorithm for our evaluations.

\begin{table}[]
\centering

\caption{Comparison between different decoding orders presented in \cref{subsec:scan}. The shown results are the transmission rate in Kilo Byte (KB) averaged over all users' requests for QP=22.}
\label{tab:different_decoding_orders}
\begin{tabular}{l|c|c|c|}
\cline{2-4}
 & \multicolumn{1}{l|}{GreedyRate} & \multicolumn{1}{l|}{SnakeLike} & \multicolumn{1}{l|}{GreedyCount} \\ \hline
\multicolumn{1}{|l|}{Market} & 1228.71 & 1217.31 & \textbf{1217.08} \\ \hline
\multicolumn{1}{|l|}{Street} & 1453.37 & 1443.25 & \textbf{1443.13} \\ \hline
\multicolumn{1}{|l|}{Mountain} & 1192.35 & 1186.69 & \textbf{1186.66} \\ \hline
\multicolumn{1}{|l|}{Church} & 1307.32 & 1299.77 & \textbf{1299.66} \\ \hline
\multicolumn{1}{|l|}{Seashore} & 1750.49 & 1745.28 & \textbf{1745.27} \\ \hline
\multicolumn{1}{|l|}{Park} & 789.16 & \textbf{781.78} & 781.79 \\ \hline
\multicolumn{1}{|l|}{Jacuzzi} & 649.85 & 645.09 & \textbf{645.06} \\ \hline
\multicolumn{1}{|l|}{Cafe} & 919.89 & 913.44 & \textbf{913.35} \\ \hline
\end{tabular}

\end{table}

\begin{figure*}[htb!]
    \centering
    \fontsize{7pt}{7pt}\selectfont
\setlength\figurewidth{0.95\linewidth}
\setlength\figureheight{0.25\linewidth}
\input{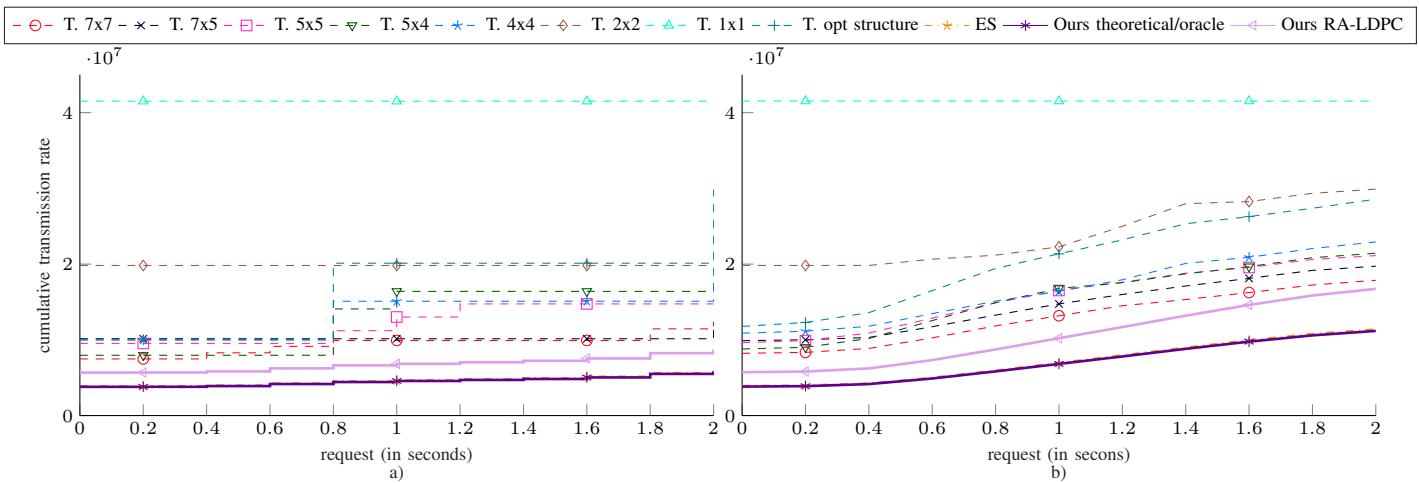}
\vspace*{-0.5cm}
    \caption{Accumulated transmission per request at iso-distortion PSNR $37$ dB for image Jacuzzi (The lower, the better). The distortion is computed at the viewport of the users. a) For one user. b) Averaged over all users' requests. ``T. $m \times n$" stands for tile-based approach where $m$ and $n$ are  the numbers of vertical and horizontal tiles and ES stands for exhaustive storage.}
	\label{fig:accumulated_transmission_per_request}
\end{figure*}

\subsection{Achievability of the oracle transmission rate}
We recall that the aim of the paper is to propose a coder that is able to achieve the \emph{oracle} transmission rate $\mathcal{R}^*$, obtained if the user head movement was known at the encoder. For that purpose, we compute the accumulated transmission rate for one user during successive requests, \emph{i.e.,}
$$R(\theta_{T'}, u)= \sum_{t=1}^{T'}r(\theta_t, u).$$
where $r(\theta_t, u)$ is the transmission rate of user $u$ with request $\theta_t$. We plot in \cref{fig:accumulated_transmission_per_request} the evolution of the accumulated transmission rate at a given PSNR of $37$dB (computed at the users' viewports) for image Jacuzzi. Similar behaviors were observed for other images and other PSNR values. 
We can see that, in theory, our proposed coder achieves the oracle transmission rate. In practice, the suboptimality of the adopted coders leaves a gap between the practical performance and the theoretical one. However, we can see that still our implementation obtains much better performance than the tile-based ones, especially at the beginning of the request. Moreover, we can observe that our method has a behavior that is better suited to interactivity than the tile-based approach. Indeed, it has a smoother $R(\theta_{T'}, u)$ evolution during the user's head navigation, meaning that our scheme transmits only what is needed upon request. By contrast, the tile-based approaches have a staircase behavior with big steps, meaning that there is a burst of rate at the moment when a new tile is needed (in which some blocks are useless). 

Another way to analyze the performance of the interactive coder is by means of the \emph{usefulness} of the transmitted data. In the following, the usefulness is defined as the proportion of displayed pixels among all the decoded ones for a request $\theta$:
$$ \mathcal{U} = \mathbb{E}_{\theta}(\mathcal{U}_\theta), $$
where
\begin{equation} \label{eq:usefulness}
\mathcal{U}_\theta = \frac{\sharp \ \mbox{displayed pixels for a request} \ \theta}{\sharp \ \mbox{decoded pixels for a request} \ \theta}.
\end{equation}
The close this value is to 1, the closer the transmitted data to what the user has requested and vice versa.

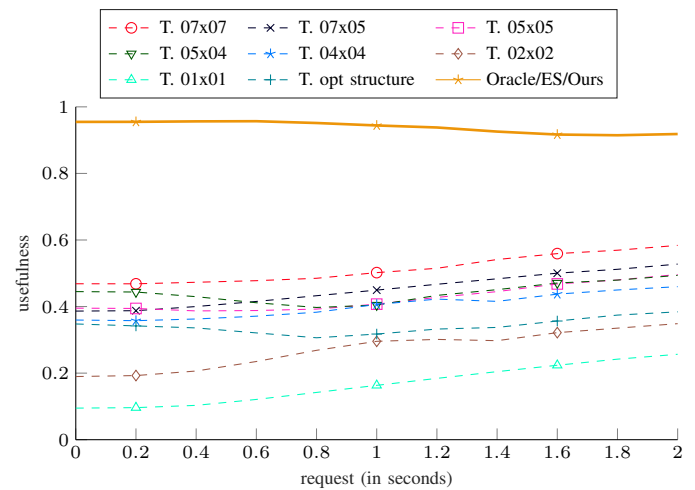
\begin{figure}[htb!]
    \centering
    \fontsize{7pt}{7pt}\selectfont
\setlength\figurewidth{0.95\linewidth}
\setlength\figureheight{0.5\linewidth}
%
%
\definecolor{mycolor1}{rgb}{1.00000,0.00000,0.10345}%
\definecolor{mycolor2}{rgb}{0.00000,0.00000,0.17241}%
\definecolor{mycolor3}{rgb}{1.00000,0.13793,0.75862}%
\definecolor{mycolor4}{rgb}{0.00000,0.34483,0.00000}%
\definecolor{mycolor5}{rgb}{0.00000,0.48276,1.00000}%
\definecolor{mycolor6}{rgb}{0.62069,0.34483,0.27586}%
\definecolor{mycolor7}{rgb}{0.00000,1.00000,0.75862}%
\definecolor{mycolor8}{rgb}{0.00000,0.51724,0.58621}%
\definecolor{mycolor9}{rgb}{0.93103,0.58621,0.03448}%
\begin{tikzpicture}

\begin{axis}[%
width=0.951\figurewidth,
height=\figureheight,
at={(0\figurewidth,0\figureheight)},
scale only axis,
xmin=0,
xmax=2,
xtick={  0, 0.2, 0.4, 0.6, 0.8,   1, 1.2, 1.4, 1.6, 1.8,   2},
xlabel style={font=\color{white!15!black}},
xlabel={request (in seconds)},
ymin=0,
ymax=1,
ylabel style={font=\color{white!15!black}},
ylabel={usefulness},
axis background/.style={fill=white},
axis x line*=bottom,
axis y line*=left,
legend style={at={(0.04,1.025)}, anchor=south west, legend columns=3, legend cell align=left, align=left, draw=white!15!black, fill=none, /tikz/every even column/.append style={column sep=0.2cm}}
]
\addplot [color=mycolor1, dashed, forget plot]
  table[row sep=crcr]{%
0	0.468493692307692\\
0.2	0.468355743589744\\
0.4	0.472976282051282\\
0.6	0.477751230769231\\
0.8	0.484963871794872\\
1	0.50196741025641\\
1.2	0.514974641025641\\
1.4	0.541321692307692\\
1.6	0.559107128205128\\
1.8	0.569259974358974\\
2	0.583826205128205\\
};
\addplot [color=mycolor1, dashed, mark=o, mark options={solid, mycolor1}]
  table[row sep=crcr]{%
1	0.50196741025641\\
};
\addlegendentry{T. 07x07}

\addplot [color=mycolor1, dashed, mark=o, mark options={solid, mycolor1}, forget plot]
  table[row sep=crcr]{%
0.2	0.468355743589744\\
};
\addplot [color=mycolor1, dashed, mark=o, mark options={solid, mycolor1}, forget plot]
  table[row sep=crcr]{%
1.6	0.559107128205128\\
};
\addplot [color=mycolor2, dashed, forget plot]
  table[row sep=crcr]{%
0	0.386388871794872\\
0.2	0.387604564102564\\
0.4	0.400082512820513\\
0.6	0.415364871794872\\
0.8	0.432596179487179\\
1	0.449449820512821\\
1.2	0.466952743589744\\
1.4	0.483413230769231\\
1.6	0.500314282051282\\
1.8	0.511941923076923\\
2	0.527593846153846\\
};
\addplot [color=mycolor2, dashed, mark=x, mark options={solid, mycolor2}]
  table[row sep=crcr]{%
1	0.449449820512821\\
};
\addlegendentry{T. 07x05}

\addplot [color=mycolor2, dashed, mark=x, mark options={solid, mycolor2}, forget plot]
  table[row sep=crcr]{%
0.2	0.387604564102564\\
};
\addplot [color=mycolor2, dashed, mark=x, mark options={solid, mycolor2}, forget plot]
  table[row sep=crcr]{%
1.6	0.500314282051282\\
};
\addplot [color=mycolor3, dashed, forget plot]
  table[row sep=crcr]{%
0	0.395213769230769\\
0.2	0.394277641025641\\
0.4	0.386934\\
0.6	0.388071435897436\\
0.8	0.392418641025641\\
1	0.40726141025641\\
1.2	0.427550307692308\\
1.4	0.444906769230769\\
1.6	0.467987666666667\\
1.8	0.479885923076923\\
2	0.496326512820513\\
};
\addplot [color=mycolor3, dashed, mark=square, mark options={solid, mycolor3}]
  table[row sep=crcr]{%
1	0.40726141025641\\
};
\addlegendentry{T. 05x05}

\addplot [color=mycolor3, dashed, mark=square, mark options={solid, mycolor3}, forget plot]
  table[row sep=crcr]{%
0.2	0.394277641025641\\
};
\addplot [color=mycolor3, dashed, mark=square, mark options={solid, mycolor3}, forget plot]
  table[row sep=crcr]{%
1.6	0.467987666666667\\
};
\addplot [color=mycolor4, dashed, forget plot]
  table[row sep=crcr]{%
0	0.444944282051282\\
0.2	0.443678461538461\\
0.4	0.429684256410256\\
0.6	0.411467282051282\\
0.8	0.397263897435898\\
1	0.404489256410256\\
1.2	0.43356141025641\\
1.4	0.450822897435897\\
1.6	0.470707538461538\\
1.8	0.479528948717949\\
2	0.493765692307692\\
};
\addplot [color=mycolor4, dashed, mark=triangle, mark options={solid, rotate=180, mycolor4}]
  table[row sep=crcr]{%
1	0.404489256410256\\
};
\addlegendentry{T. 05x04}

\addplot [color=mycolor4, dashed, mark=triangle, mark options={solid, rotate=180, mycolor4}, forget plot]
  table[row sep=crcr]{%
0.2	0.443678461538461\\
};
\addplot [color=mycolor4, dashed, mark=triangle, mark options={solid, rotate=180, mycolor4}, forget plot]
  table[row sep=crcr]{%
1.6	0.470707538461538\\
};
\addplot [color=mycolor5, dashed, forget plot]
  table[row sep=crcr]{%
0	0.359504948717949\\
0.2	0.357598102564103\\
0.4	0.362681641025641\\
0.6	0.371145205128205\\
0.8	0.383017512820513\\
1	0.406296282051282\\
1.2	0.422456102564102\\
1.4	0.415580717948718\\
1.6	0.437342846153846\\
1.8	0.449838846153846\\
2	0.459773692307692\\
};
\addplot [color=mycolor5, dashed, mark=star, mark options={solid, mycolor5}]
  table[row sep=crcr]{%
1	0.406296282051282\\
};
\addlegendentry{T. 04x04}

\addplot [color=mycolor5, dashed, mark=star, mark options={solid, mycolor5}, forget plot]
  table[row sep=crcr]{%
0.2	0.357598102564103\\
};
\addplot [color=mycolor5, dashed, mark=star, mark options={solid, mycolor5}, forget plot]
  table[row sep=crcr]{%
1.6	0.437342846153846\\
};
\addplot [color=mycolor6, dashed, forget plot]
  table[row sep=crcr]{%
0	0.189737333333333\\
0.2	0.192707846153846\\
0.4	0.206335333333333\\
0.6	0.234934025641026\\
0.8	0.26852241025641\\
1	0.295580461538462\\
1.2	0.301306\\
1.4	0.297525384615385\\
1.6	0.321620461538461\\
1.8	0.335153615384615\\
2	0.348629205128205\\
};
\addplot [color=mycolor6, dashed, mark=diamond, mark options={solid, mycolor6}]
  table[row sep=crcr]{%
1	0.295580461538462\\
};
\addlegendentry{T. 02x02}

\addplot [color=mycolor6, dashed, mark=diamond, mark options={solid, mycolor6}, forget plot]
  table[row sep=crcr]{%
0.2	0.192707846153846\\
};
\addplot [color=mycolor6, dashed, mark=diamond, mark options={solid, mycolor6}, forget plot]
  table[row sep=crcr]{%
1.6	0.321620461538461\\
};
\addplot [color=mycolor7, dashed, forget plot]
  table[row sep=crcr]{%
0	0.0948686923076923\\
0.2	0.0963539230769231\\
0.4	0.103167764102564\\
0.6	0.120966928205128\\
0.8	0.142028664102564\\
1	0.163293987179487\\
1.2	0.18423208974359\\
1.4	0.204489666666667\\
1.6	0.223666641025641\\
1.8	0.24195682051282\\
2	0.256631615384615\\
};
\addplot [color=mycolor7, dashed, mark=triangle, mark options={solid, mycolor7}]
  table[row sep=crcr]{%
1	0.163293987179487\\
};
\addlegendentry{T. 01x01}

\addplot [color=mycolor7, dashed, mark=triangle, mark options={solid, mycolor7}, forget plot]
  table[row sep=crcr]{%
0.2	0.0963539230769231\\
};
\addplot [color=mycolor7, dashed, mark=triangle, mark options={solid, mycolor7}, forget plot]
  table[row sep=crcr]{%
1.6	0.223666641025641\\
};
\addplot [color=mycolor8, dashed, forget plot]
  table[row sep=crcr]{%
0	0.347704384615385\\
0.2	0.341883230769231\\
0.4	0.335855384615385\\
0.6	0.320846666666667\\
0.8	0.306236743589744\\
1	0.317419102564102\\
1.2	0.332395564102564\\
1.4	0.337323923076923\\
1.6	0.356587692307692\\
1.8	0.374363461538461\\
2	0.384154282051282\\
};
\addplot [color=mycolor8, dashed, mark=+, mark options={solid, mycolor8}]
  table[row sep=crcr]{%
1	0.317419102564102\\
};
\addlegendentry{T. opt structure}

\addplot [color=mycolor8, dashed, mark=+, mark options={solid, mycolor8}, forget plot]
  table[row sep=crcr]{%
0.2	0.341883230769231\\
};
\addplot [color=mycolor8, dashed, mark=+, mark options={solid, mycolor8}, forget plot]
  table[row sep=crcr]{%
1.6	0.356587692307692\\
};
\addplot [color=mycolor9, line width=1.0pt, forget plot]
  table[row sep=crcr]{%
0	0.954815948717949\\
0.2	0.954981256410256\\
0.4	0.956390615384616\\
0.6	0.956865102564103\\
0.8	0.951478820512821\\
1	0.944114076923077\\
1.2	0.938010794871795\\
1.4	0.925701384615385\\
1.6	0.916821435897436\\
1.8	0.914660102564103\\
2	0.918376102564103\\
};
\addplot [color=mycolor9, mark=star, mark options={solid, mycolor9}]
  table[row sep=crcr]{%
1	0.944114076923077\\
};
\addlegendentry{Oracle/ES/Ours}

\addplot [color=mycolor9, mark=star, mark options={solid, mycolor9}, forget plot]
  table[row sep=crcr]{%
0.2	0.954981256410256\\
};
\addplot [color=mycolor9, mark=star, mark options={solid, mycolor9}, forget plot]
  table[row sep=crcr]{%
1.6	0.916821435897436\\
};
\end{axis}
\end{tikzpicture}%
    \caption{Usefulness of the requested pixels per request average of all users' requests for image Jacuzzi. ``T. $m \times n$" stands for tile-based approach where $m$ and $n$ are  the numbers of vertical and horizontal tiles and ES stands for exhaustive storage.}
	\label{fig:usefulness}
\end{figure}

The usefulness averaged over all users is shown in \cref{fig:usefulness} for image Jacuzzi. It confirms that our scheme sends more useful information at each request compared to its baselines.

In all these experiments, the ES scheme performs as good as our solution in terms of transmission rate. More exactly, it performs as good as our theoretical performance when using an optimal rate-adaptive code. In practice, they use an arithmetic coder that does not suffer from some sub-optimality of rate-adaptive implementation. However, in the ES approach, the focus is on the transmission rate only, and this is why in the following we present an evaluation that includes the storage cost.


\subsection{Analysis of the Rate-storage trade-off}
We now discuss the performance of the different coders in terms of rate and storage performance. For that, we use a weighted Bjontegaard metric introduced in \cite{mahmoudianicip2019}. In this metric the weight $\lambda$ balances the relative importance between the rate and the storage, as a function of the application:
\begin{equation} \label{eq:linear_combination_storage_transmission_lambda}
\mathbb{E}_{\theta}(R_\theta) + \lambda S.
\end{equation} 
 In the following, we first study two extreme scenarios where either the rate or the storage is neglected. Then, we show results for more realistic scenarios where both matters.

\subsubsection{RD and SD performances}
\begin{figure*}[htb!]
    \centering
    \fontsize{7pt}{7pt}\selectfont
\setlength\figurewidth{0.95\linewidth}
\setlength\figureheight{0.25\linewidth}
%
%
\definecolor{mycolor1}{rgb}{1.00000,0.00000,0.10345}%
\definecolor{mycolor2}{rgb}{0.00000,0.00000,0.17241}%
\definecolor{mycolor3}{rgb}{1.00000,0.13793,0.75862}%
\definecolor{mycolor4}{rgb}{0.00000,0.34483,0.00000}%
\definecolor{mycolor5}{rgb}{0.00000,0.48276,1.00000}%
\definecolor{mycolor6}{rgb}{0.62069,0.34483,0.27586}%
\definecolor{mycolor7}{rgb}{0.00000,1.00000,0.75862}%
\definecolor{mycolor8}{rgb}{0.00000,0.51724,0.58621}%
\definecolor{mycolor9}{rgb}{0.93103,0.58621,0.03448}%
\definecolor{mycolor10}{rgb}{0.37931,0.00000,0.55172}%
\definecolor{mycolor11}{rgb}{0.86207,0.62069,0.93103}%
\begin{tikzpicture}

\begin{axis}[%
width=0.489\figurewidth,
height=\figureheight,
at={(0\figurewidth,0\figureheight)},
scale only axis,
xmin=1000000,
xmax=10000000,
xlabel style={font=\color{white!15!black}},
xlabel={${E}_{\theta}(R_\theta)$ (in bits)},
ymin=0.86,
ymax=0.98,
ylabel style={font=\color{white!15!black}},
ylabel={SSIM},
axis background/.style={fill=white},
axis x line*=bottom,
axis y line*=left
]
\addplot [color=mycolor1, dashed, mark=o, mark options={solid, mycolor1}, forget plot]
  table[row sep=crcr]{%
3141735.96581197	0.975064320512821\\
2658648.03418803	0.950207376068376\\
2353219.62393162	0.918938581196581\\
2197851.41880342	0.890770700854701\\
2108669.53846154	0.860190405982906\\
};
\addplot [color=mycolor2, dashed, mark=x, mark options={solid, mycolor2}, forget plot]
  table[row sep=crcr]{%
3487106.32478632	0.975060756410256\\
2958047.07692308	0.950298196581197\\
2624192.78632479	0.918944081196581\\
2458099.14529914	0.891172688034188\\
2364627.21367521	0.860293299145299\\
};
\addplot [color=mycolor3, dashed, mark=square, mark options={solid, mycolor3}, forget plot]
  table[row sep=crcr]{%
3954640.64957265	0.975039226495727\\
3349661.47008547	0.950232290598291\\
2958432.23931624	0.91880597008547\\
2753655.41880342	0.890765239316239\\
2632778.29059829	0.860841692307692\\
};
\addplot [color=mycolor4, dashed, mark=triangle, mark options={solid, rotate=180, mycolor4}, forget plot]
  table[row sep=crcr]{%
4027652.20512821	0.975062358974359\\
3410406.73504274	0.950234679487179\\
3010416.78632479	0.918639717948718\\
2798930.18803419	0.891356833333333\\
2673915.52136752	0.86097511965812\\
};
\addplot [color=mycolor5, dashed, mark=star, mark options={solid, mycolor5}, forget plot]
  table[row sep=crcr]{%
3892882.83760684	0.975062782051282\\
3305253.70940171	0.95028\\
2929392.23931624	0.918550948717949\\
2738611.14529914	0.891040008547008\\
2628875.31623932	0.860276602564103\\
};
\addplot [color=mycolor6, dashed, mark=diamond, mark options={solid, mycolor6}, forget plot]
  table[row sep=crcr]{%
5055975.76068376	0.975060517094017\\
4370169.94871795	0.950243106837607\\
3938765.77777778	0.918670324786325\\
3720220.27350427	0.890903568376068\\
3595307.31623931	0.86038558974359\\
};
\addplot [color=mycolor7, dashed, mark=triangle, mark options={solid, mycolor7}, forget plot]
  table[row sep=crcr]{%
9728338.66666667	0.975055833333333\\
8379066.66666667	0.950219461538462\\
7471422.66666667	0.91884697008547\\
6947530.66666666	0.891153568376069\\
6612942.66666666	0.860594264957265\\
};
\addplot [color=mycolor8, dashed, mark=+, mark options={solid, mycolor8}, forget plot]
  table[row sep=crcr]{%
5068785.67521368	0.975053341880342\\
4301624.95726496	0.950261602564102\\
3809440.1025641	0.918585217948718\\
3562335.21367521	0.891248397435897\\
3419948.82051282	0.860491961538462\\
};
\addplot [color=mycolor9, dashdotted, mark=star, mark options={solid, mycolor9}, forget plot]
  table[row sep=crcr]{%
1623817.64102564	0.975367833333334\\
1372894.52991453	0.951367239316239\\
1224794.42735043	0.921486260683761\\
1156870.11965812	0.895584282051282\\
1119686.90598291	0.867555987179487\\
};
\addplot [color=mycolor10, mark=asterisk, mark options={solid, mycolor10}, forget plot]
  table[row sep=crcr]{%
1604855.06410256	0.975368183760684\\
1354034.91452991	0.951367423076923\\
1206223.57264957	0.921486713675214\\
1138497.15811966	0.895584572649573\\
1101558.02564103	0.867556623931624\\
};
\addplot [color=mycolor11, mark=triangle, mark options={solid, rotate=90, mycolor11}, forget plot]
  table[row sep=crcr]{%
2403677.43162393	0.975368183760684\\
2027539.53632479	0.951367423076923\\
1805895.64316239	0.921486713675214\\
1704359.7991453	0.895584572649573\\
1649004.90384615	0.867556623931624\\
};
\end{axis}

\begin{axis}[%
width=0.489\figurewidth,
height=\figureheight,
at={(0.511\figurewidth,0\figureheight)},
scale only axis,
xmin=0,
xmax=700000000,
xlabel style={font=\color{white!15!black}},
xlabel={$S$ (in bits)},
ymin=0.86,
ymax=0.98,
ylabel style={font=\color{white!15!black}},
ylabel={SSIM},
axis background/.style={fill=white},
axis x line*=bottom,
axis y line*=left,
legend style={at={(-0.05,1.2)}, anchor=north, legend columns=11, legend cell align=left, align=left, draw=white!15!black}
]
\addplot [color=mycolor1, dashed, mark=o, mark options={solid, mycolor1}]
  table[row sep=crcr]{%
58389944	0.975064320512821\\
50300168	0.950207376068376\\
44854128	0.918938581196581\\
41708664	0.890770700854701\\
39702344	0.860190405982906\\
};
\addlegendentry{T. 7x7}

\addplot [color=mycolor2, dashed, mark=x, mark options={solid, mycolor2}]
  table[row sep=crcr]{%
58386928	0.975060756410256\\
50294768	0.950298196581197\\
44847088	0.918944081196581\\
41702816	0.891172688034188\\
39696816	0.860293299145299\\
};
\addlegendentry{T. 7x5}

\addplot [color=mycolor3, dashed, mark=square, mark options={solid, mycolor3}]
  table[row sep=crcr]{%
58381760	0.975039226495727\\
50290576	0.950232290598291\\
44841008	0.91880597008547\\
41701072	0.890765239316239\\
39695200	0.860841692307692\\
};
\addlegendentry{T. 5x5}

\addplot [color=mycolor4, dashed, mark=triangle, mark options={solid, rotate=180, mycolor4}]
  table[row sep=crcr]{%
58377400	0.975062358974359\\
50286536	0.950234679487179\\
44842784	0.918639717948718\\
41698728	0.891356833333333\\
39689280	0.86097511965812\\
};
\addlegendentry{T. 5x4}

\addplot [color=mycolor5, dashed, mark=star, mark options={solid, mycolor5}]
  table[row sep=crcr]{%
58378152	0.975062782051282\\
50286568	0.95028\\
44837768	0.918550948717949\\
41699640	0.891040008547008\\
39690824	0.860276602564103\\
};
\addlegendentry{T. 4x4}

\addplot [color=mycolor6, dashed, mark=diamond, mark options={solid, mycolor6}]
  table[row sep=crcr]{%
58374336	0.975060517094017\\
50281936	0.950243106837607\\
44834600	0.918670324786325\\
41692416	0.890903568376068\\
39685392	0.86038558974359\\
};
\addlegendentry{T. 2x2}

\addplot [color=mycolor7, dashed, mark=triangle, mark options={solid, mycolor7}]
  table[row sep=crcr]{%
58370032	0.975055833333333\\
50274400	0.950219461538462\\
44828536	0.91884697008547\\
41685184	0.891153568376069\\
39677656	0.860594264957265\\
};
\addlegendentry{T. 1x1}

\addplot [color=mycolor8, dashed, mark=+, mark options={solid, mycolor8}]
  table[row sep=crcr]{%
58371832	0.975053341880342\\
50279224	0.950261602564102\\
44832096	0.918585217948718\\
41693600	0.891248397435897\\
39685640	0.860491961538462\\
};
\addlegendentry{T. opt structure}

\addplot [color=mycolor9, dashdotted, mark=star, mark options={solid, mycolor9}]
  table[row sep=crcr]{%
698009792	0.975367833333334\\
602090464	0.951367239316239\\
536646392	0.921486260683761\\
498804296	0.895584282051282\\
474221232	0.867555987179487\\
};
\addlegendentry{ES}

\addplot [color=mycolor10, mark=asterisk, mark options={solid, mycolor10}]
  table[row sep=crcr]{%
61055742	0.975368183760684\\
52973593	0.951367423076923\\
47480243	0.921486713675214\\
44265549	0.895584572649573\\
42122410	0.867556623931624\\
};
\addlegendentry{Ours theoretical}

\addplot [color=mycolor11, mark=triangle, mark options={solid, rotate=90, mycolor11}]
  table[row sep=crcr]{%
90153750	0.975368183760684\\
78066049.5	0.951367423076923\\
69854669.5	0.921486713675214\\
65056343.5	0.895584572649573\\
61861826	0.867556623931624\\
};
\addlegendentry{Ours RA-LDPC}

\end{axis}
\end{tikzpicture}%
    \caption{SSIM of luma channel versus storage and transmission rate averaged after 1 sec of users' requests for image Jacuzzi. a) Transmission rate versus SSIM curve. b) Storage vs SSIM curve. ``T. $m \times n$" stands for tile-based approach where $m$ and $n$ are  the numbers of vertical and horizontal tiles and ES stands for exhaustive storage.}
	\label{fig:rate_storage_SSIM_curve}
\end{figure*}
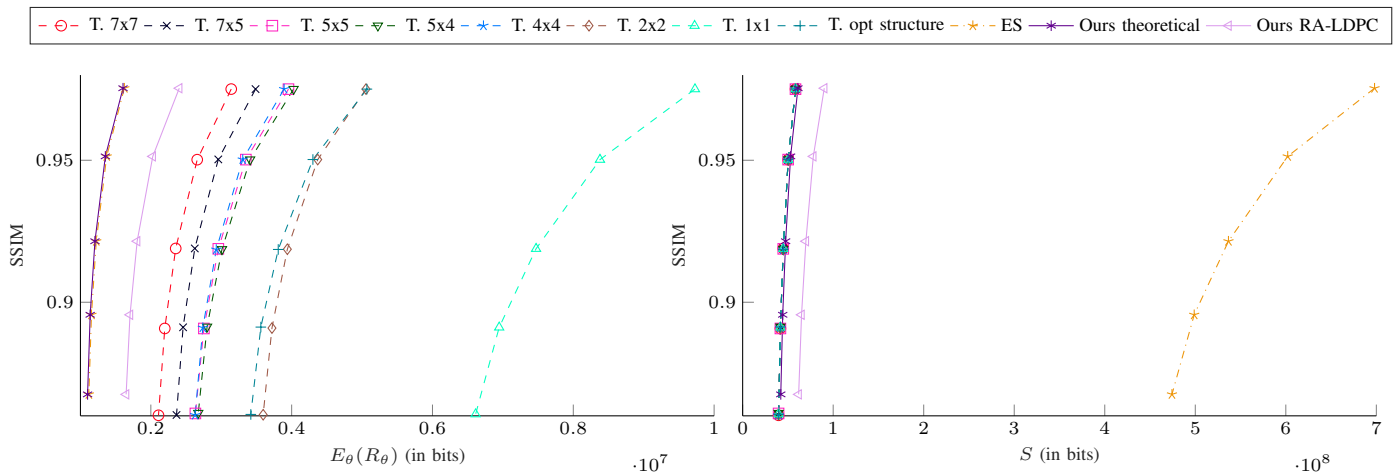

In \cref{fig:RD_SD_curves}, for queries of length \emph{1 sec} averaged over all users, we compare the performance when either the storage ($\lambda =  0$) or the transmission rate ($\lambda=\infty$) is neglected. We see that the proposed method with RA-LDPC codes performs better than tiling approaches with only a small extra cost in storage. Compared to the ES approach, the ES performs better on the transmission rate, which is basically because rate-adaptive channel codes are not optimal for short-length sequences, but as expected, the ES approach performs poorly in terms of storage. We also computed the SSIM value of the user's viewport luma channel. The same conclusion as storage/transmission rate-distortion holds also for SSIM value. The curves for transmission and storage versus SSIM values are shown in \cref{fig:rate_storage_SSIM_curve}. Similar behaviors observed for other images as well.

\begin{figure*}[htb!]
    \centering
    \fontsize{7pt}{7pt}\selectfont
\setlength\figurewidth{0.95\linewidth}
\setlength\figureheight{0.25\linewidth}
%
%
\definecolor{mycolor1}{rgb}{1.00000,0.00000,0.10345}%
\definecolor{mycolor2}{rgb}{0.00000,0.00000,0.17241}%
\definecolor{mycolor3}{rgb}{1.00000,0.13793,0.75862}%
\definecolor{mycolor4}{rgb}{0.00000,0.34483,0.00000}%
\definecolor{mycolor5}{rgb}{0.00000,0.48276,1.00000}%
\definecolor{mycolor6}{rgb}{0.62069,0.34483,0.27586}%
\definecolor{mycolor7}{rgb}{0.00000,1.00000,0.75862}%
\definecolor{mycolor8}{rgb}{0.00000,0.51724,0.58621}%
\definecolor{mycolor9}{rgb}{0.93103,0.58621,0.03448}%
\definecolor{mycolor10}{rgb}{0.37931,0.00000,0.55172}%
\definecolor{mycolor11}{rgb}{0.86207,0.62069,0.93103}%
\begin{tikzpicture}

\begin{axis}[%
width=0.489\figurewidth,
height=\figureheight,
at={(0\figurewidth,0\figureheight)},
scale only axis,
xmin=1000000,
xmax=10000000,
xlabel style={font=\color{white!15!black}},
xlabel={${E}_{\theta}(R_\theta)$ (in bits)},
ymin=34,
ymax=44,
ylabel style={font=\color{white!15!black}},
ylabel={PSNR},
axis background/.style={fill=white},
axis x line*=bottom,
axis y line*=left
]
\addplot [color=mycolor1, dashed, mark=o, mark options={solid, mycolor1}, forget plot]
  table[row sep=crcr]{%
3141735.96581197	43.2362523410475\\
2658648.03418803	41.0458948835022\\
2353219.62393162	38.9197348464803\\
2197851.41880342	37.0269214802206\\
2108669.53846154	34.968249009715\\
};
\addplot [color=mycolor2, dashed, mark=x, mark options={solid, mycolor2}, forget plot]
  table[row sep=crcr]{%
3487106.32478632	43.2368520293547\\
2958047.07692308	41.0554885958949\\
2624192.78632479	38.9191483096727\\
2458099.14529914	37.0375899887042\\
2364627.21367521	34.9579951758447\\
};
\addplot [color=mycolor3, dashed, mark=square, mark options={solid, mycolor3}, forget plot]
  table[row sep=crcr]{%
3954640.64957265	43.2330038452793\\
3349661.47008547	41.047462781848\\
2958432.23931624	38.921792858472\\
2753655.41880342	37.0272128237093\\
2632778.29059829	34.9748464139242\\
};
\addplot [color=mycolor4, dashed, mark=triangle, mark options={solid, rotate=180, mycolor4}, forget plot]
  table[row sep=crcr]{%
4027652.20512821	43.2380482254437\\
3410406.73504274	41.0492516573316\\
3010416.78632479	38.9124208999715\\
2798930.18803419	37.0376982102583\\
2673915.52136752	34.9583698243495\\
};
\addplot [color=mycolor5, dashed, mark=star, mark options={solid, mycolor5}, forget plot]
  table[row sep=crcr]{%
3892882.83760684	43.2396513091208\\
3305253.70940171	41.0482675447459\\
2929392.23931624	38.9109966548331\\
2738611.14529914	37.0366003916041\\
2628875.31623932	34.9411500339416\\
};
\addplot [color=mycolor6, dashed, mark=diamond, mark options={solid, mycolor6}, forget plot]
  table[row sep=crcr]{%
5055975.76068376	43.2396419812084\\
4370169.94871795	41.0497128301633\\
3938765.77777778	38.9133539882015\\
3720220.27350427	37.0293769198786\\
3595307.31623931	34.9545624477022\\
};
\addplot [color=mycolor7, dashed, mark=triangle, mark options={solid, mycolor7}, forget plot]
  table[row sep=crcr]{%
9728338.66666667	43.2400646460199\\
8379066.66666667	41.0526548543708\\
7471422.66666667	38.9153925911138\\
6947530.66666666	37.0216329067394\\
6612942.66666666	34.9555332761819\\
};
\addplot [color=mycolor8, dashed, mark=+, mark options={solid, mycolor8}, forget plot]
  table[row sep=crcr]{%
5068785.67521368	43.2404780824535\\
4301624.95726496	41.0508289845856\\
3809440.1025641	38.9149660209619\\
3562335.21367521	37.037567190374\\
3419948.82051282	34.934351708233\\
};
\addplot [color=mycolor9, dashdotted, mark=star, mark options={solid, mycolor9}, forget plot]
  table[row sep=crcr]{%
1623817.64102564	43.1807124877121\\
1372894.52991453	41.020316811884\\
1224794.42735043	38.9344955565297\\
1156870.11965812	37.1152835328397\\
1119686.90598291	35.1597420115427\\
};
\addplot [color=mycolor10, mark=asterisk, mark options={solid, mycolor10}, forget plot]
  table[row sep=crcr]{%
1604855.06410256	43.1807114190712\\
1354034.91452991	41.020318219825\\
1206223.57264957	38.9344977227843\\
1138497.15811966	37.1152850018219\\
1101558.02564103	35.159742667024\\
};
\addplot [color=mycolor11, mark=triangle, mark options={solid, rotate=90, mycolor11}, forget plot]
  table[row sep=crcr]{%
2403677.43162393	43.1807114190712\\
2027539.53632479	41.020318219825\\
1805895.64316239	38.9344977227843\\
1704359.7991453	37.1152850018219\\
1649004.90384615	35.159742667024\\
};
\end{axis}

\begin{axis}[%
width=0.489\figurewidth,
height=\figureheight,
at={(0.511\figurewidth,0\figureheight)},
scale only axis,
xmin=0,
xmax=700000000,
xlabel style={font=\color{white!15!black}},
xlabel={$S$ (in bits)},
ymin=34,
ymax=44,
ylabel style={font=\color{white!15!black}},
ylabel={PSNR},
axis background/.style={fill=white},
axis x line*=bottom,
axis y line*=left,
legend style={at={(-0.05,1.2)}, anchor=north, legend columns=11, legend cell align=left, align=left, draw=white!15!black}
]
\addplot [color=mycolor1, dashed, mark=o, mark options={solid, mycolor1}]
  table[row sep=crcr]{%
58389944	43.2362523410475\\
50300168	41.0458948835022\\
44854128	38.9197348464803\\
41708664	37.0269214802206\\
39702344	34.968249009715\\
};
\addlegendentry{T. 7x7}

\addplot [color=mycolor2, dashed, mark=x, mark options={solid, mycolor2}]
  table[row sep=crcr]{%
58386928	43.2368520293547\\
50294768	41.0554885958949\\
44847088	38.9191483096727\\
41702816	37.0375899887042\\
39696816	34.9579951758447\\
};
\addlegendentry{T. 7x5}

\addplot [color=mycolor3, dashed, mark=square, mark options={solid, mycolor3}]
  table[row sep=crcr]{%
58381760	43.2330038452793\\
50290576	41.047462781848\\
44841008	38.921792858472\\
41701072	37.0272128237093\\
39695200	34.9748464139242\\
};
\addlegendentry{T. 5x5}

\addplot [color=mycolor4, dashed, mark=triangle, mark options={solid, rotate=180, mycolor4}]
  table[row sep=crcr]{%
58377400	43.2380482254437\\
50286536	41.0492516573316\\
44842784	38.9124208999715\\
41698728	37.0376982102583\\
39689280	34.9583698243495\\
};
\addlegendentry{T. 5x4}

\addplot [color=mycolor5, dashed, mark=star, mark options={solid, mycolor5}]
  table[row sep=crcr]{%
58378152	43.2396513091208\\
50286568	41.0482675447459\\
44837768	38.9109966548331\\
41699640	37.0366003916041\\
39690824	34.9411500339416\\
};
\addlegendentry{T. 4x4}

\addplot [color=mycolor6, dashed, mark=diamond, mark options={solid, mycolor6}]
  table[row sep=crcr]{%
58374336	43.2396419812084\\
50281936	41.0497128301633\\
44834600	38.9133539882015\\
41692416	37.0293769198786\\
39685392	34.9545624477022\\
};
\addlegendentry{T. 2x2}

\addplot [color=mycolor7, dashed, mark=triangle, mark options={solid, mycolor7}]
  table[row sep=crcr]{%
58370032	43.2400646460199\\
50274400	41.0526548543708\\
44828536	38.9153925911138\\
41685184	37.0216329067394\\
39677656	34.9555332761819\\
};
\addlegendentry{T. 1x1}

\addplot [color=mycolor8, dashed, mark=+, mark options={solid, mycolor8}]
  table[row sep=crcr]{%
58371832	43.2404780824535\\
50279224	41.0508289845856\\
44832096	38.9149660209619\\
41693600	37.037567190374\\
39685640	34.934351708233\\
};
\addlegendentry{T. opt structure}

\addplot [color=mycolor9, dashdotted, mark=star, mark options={solid, mycolor9}]
  table[row sep=crcr]{%
698009792	43.1807124877121\\
602090464	41.020316811884\\
536646392	38.9344955565297\\
498804296	37.1152835328397\\
474221232	35.1597420115427\\
};
\addlegendentry{ES}

\addplot [color=mycolor10, mark=asterisk, mark options={solid, mycolor10}]
  table[row sep=crcr]{%
61055742	43.1807114190712\\
52973593	41.020318219825\\
47480243	38.9344977227843\\
44265549	37.1152850018219\\
42122410	35.159742667024\\
};
\addlegendentry{Ours theoretical}

\addplot [color=mycolor11, mark=triangle, mark options={solid, rotate=90, mycolor11}]
  table[row sep=crcr]{%
90153750	43.1807114190712\\
78066049.5	41.020318219825\\
69854669.5	38.9344977227843\\
65056343.5	37.1152850018219\\
61861826	35.159742667024\\
};
\addlegendentry{Ours RA-LDPC}

\end{axis}
\end{tikzpicture}%
    \caption{Storage and transmission performance after 1 sec of users' requests for image Jacuzzi. a) Transmission rate-distortion curve. b) Storage-distortion curve. ``T. $m \times n$" stands for tile-based approach where $m$ and $n$ are  the numbers of vertical and horizontal tiles and ES stands for exhaustive storage.}
	\label{fig:RD_SD_curves}
\end{figure*}
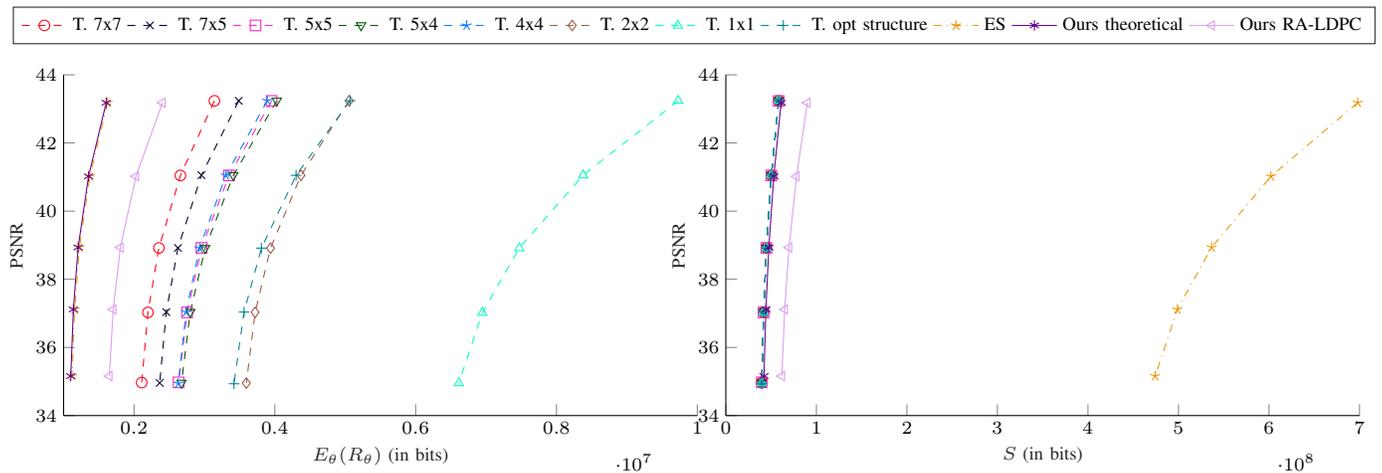

The averaged bitrate differences in transmission rate and storage are summarized in \cref{tab:BD-R_BD-S} for different length of users' requests. The bitrate difference show the performance of each method relative to the no tiling (T. $1\times 1$) approach, i.e., when the whole image is stored and transmitted at once. Negative values indicate bitrate saving w.r.t. no tiling approach. Due to space preservation, intermediary tile sizes are removed from the table and only the important baselines are shown. It is observed that while the performance of RA-LDPC codes does not exactly reach its theoretical rates,  they still perform better than the tile-based approaches in terms of transmission rate even at a long duration of 2 secs of user's navigation in a single image (it is worth noting that usually after almost 4 secs, users are viewing the whole image). In terms of storage performance, we observe that the proposed method performs significantly better than ES (the greater the positive value of BD, the worse the performance). In other words, the proposed method loses $56.36\%$ of storage w.r.t no tiling approach while exhaustive storage loses $1102.31\%$. If we look at the theoretical results of the proposed coder, we see that in theory this coder should only lose $6.14\%$ in terms of storage and gain $89.72\%$ in transmission rate for the first request of the user, which clearly shows that there is plenty of room for improvement in designing rate-adaptive channel codes.

\begin{table*}[htb!]
\centering
\caption{BD measures (in percent) averaged over all users relative to the no tiling approach (T. 1x1). BD-R $n$ is the BD metric for transmission rate for requests of length $n$. BD-S is the BD-storage measure.}
\label{tab:BD-R_BD-S}
\begin{tabular}{cccccccccc||c}
\hline
\multicolumn{2}{c}{} & Market & Street & Mountain & Church & Seashore & Park & Jacuzzi & Cafe & Average \\ \hline
\multicolumn{1}{c|}{\multirow{5}{*}{\begin{tabular}[c]{@{}c@{}}BD-R \\ 1st\\ request\end{tabular}}} & T. 2x2 & -50.66 & -51.65 & -49.74 & -50.96 & -46.56 & -51.19 & -53.13 & -49.03 & -50.36 \\ \cline{2-11} 
\multicolumn{1}{c|}{} & T. 7x7 & -79.77 & -78.39 & -78.25 & -79.67 & -74.96 & -79.34 & -80.34 & -77.73 & -78.56 \\ \cline{2-11} 
\multicolumn{1}{c|}{} & ES & -90.39 & -89.36 & -89.77 & -89.59 & -87.68 & -90.22 & -90.80 & -88.93 & -89.59 \\ \cline{2-11} 
\multicolumn{1}{c|}{} & Ours theoretical & -90.52 & -89.49 & -89.91 & -89.71 & -87.81 & -90.36 & -90.94 & -89.06 & -89.72 \\ \cline{2-11} 
\multicolumn{1}{c|}{} & Ours RA-LDPC & -85.80 & -84.26 & -84.89 & -84.59 & -81.74 & -85.56 & -86.43 & -83.62 & -84.61 \\ \hline \hline
\multicolumn{1}{c|}{\multirow{5}{*}{\begin{tabular}[c]{@{}c@{}}BD-R\\ 1 sec\end{tabular}}} & T. 2x2 & -44.38 & -39.25 & -31.14 & -48.04 & -24.32 & -44.52 & -47.15 & -39.73 & -39.82 \\ \cline{2-11} 
\multicolumn{1}{c|}{} & T. 7x7 & -73.90 & -73.39 & -71.29 & -73.39 & -68.07 & -73.07 & -68.28 & -72.32 & -71.71 \\ \cline{2-11} 
\multicolumn{1}{c|}{} & ES & -87.29 & -86.36 & -84.38 & -85.91 & -82.60 & -86.07 & -83.47 & -85.38 & -85.18 \\ \cline{2-11} 
\multicolumn{1}{c|}{} & Ours theoretical & -87.46 & -86.52 & -84.59 & -86.08 & -82.79 & -86.27 & -83.71 & -85.56 & -85.37 \\ \cline{2-11} 
\multicolumn{1}{c|}{} & Ours RA-LDPC & -81.23 & -79.82 & -76.91 & -79.15 & -74.22 & -79.45 & -75.61 & -78.38 & -78.09 \\ \hline \hline
\multicolumn{1}{c|}{\multirow{5}{*}{\begin{tabular}[c]{@{}c@{}}BD-R\\ 2 secs\end{tabular}}} & T. 2x2 & -28.41 & -23.33 & -9.40 & -31.79 & -10.63 & -23.43 & -28.95 & -19.04 & -21.87 \\ \cline{2-11} 
\multicolumn{1}{c|}{} & T. 7x7 & -69.91 & -67.69 & -64.30 & -63.98 & -59.89 & -65.50 & -57.19 & -66.42 & -64.36 \\ \cline{2-11} 
\multicolumn{1}{c|}{} & ES & -84.58 & -82.16 & -79.28 & -78.50 & -77.00 & -80.93 & -72.97 & -81.17 & -79.57 \\ \cline{2-11} 
\multicolumn{1}{c|}{} & Ours theoretical & -84.78 & -82.38 & -79.55 & -78.76 & -77.26 & -81.20 & -73.36 & -81.41 & -79.84 \\ \cline{2-11} 
\multicolumn{1}{c|}{} & Ours RA-LDPC & -77.21 & -73.61 & -69.38 & -68.19 & -65.94 & -71.85 & -60.12 & -72.16 & -69.81 \\ \hline \hline
\multicolumn{1}{c|}{\multirow{5}{*}{BD-S}} & T. 2x2 & 0.01 & 0.02 & -0.00 & 0.01 & 0.01 & 0.02 & 0.02 & 0.03 & 0.01 \\ \cline{2-11} 
\multicolumn{1}{c|}{} & T. 7x7 & 0.08 & 0.07 & 0.02 & 0.08 & 0.03 & 0.06 & 0.06 & 0.09 & 0.06 \\ \cline{2-11} 
\multicolumn{1}{c|}{} & ES & 1110.57 & 1105.01 & 1096.83 & 1103.01 & 1095.92 & 1106.26 & 1096.71 & 1104.18 & 1102.31 \\ \cline{2-11} 
\multicolumn{1}{c|}{} & Ours theoretical & 7.14 & 6.02 & 5.36 & 6.11 & 4.99 & 7.06 & 5.71 & 6.72 & 6.14 \\ \cline{2-11} 
\multicolumn{1}{c|}{} & Ours RA-LDPC & 57.86 & 56.35 & 55.15 & 56.48 & 54.75 & 57.55 & 55.58 & 57.14 & 56.36 \\ \hline
\end{tabular}
\end{table*}

\subsubsection{Weighted BD}
The weighted BD is presented for more realistic scenarios, where both storage and transmission rates matter with a relative importance.
Lower values of $\lambda$ given in \eqref{eq:linear_combination_storage_transmission_lambda} means that the importance is given more to transmission rate and, on the contrary, higher $\lambda$ values give the importance to storage. Weighted BD values of different baselines are summarized in \cref{tab:weighted_BD}. In our experiments, for tile-based and our approach, the storage is $10-100$ times larger than the transmission rate and for the exhaustive storage, the storage is usually $800-1000$ times the transmission rate. Therefore, imposing $\lambda$ around $0.01$ gives the same importance between transmission rate and storage. It can be seen from \cref{tab:weighted_BD} that for $\lambda= 0.01$ the RA-LDPC coder performs better than other baselines which means that if the storage and transmission rates are of the same importance, the proposed method performs better. Decreasing $\lambda$ to $1e^{-3}$ puts emphasis on the transmission rate where exhaustive storage is more suitable, but it can be seen that even in this case the performance of the our RA-LDPC coder is pretty close to ES coder. Therefore, in terms of transmission rate both our solution and the exhaustive storage approach have the same performance, but our coder has much lower storage overhead. Finally, if the storage is more important than the transmission rate ($\lambda \geq 0.1$) tiling approaches perform better.

\begin{table*}[htb]
\centering
\caption{Weighted BD for requests of length 1 sec averaged over all users relative to the no tiling approach (T. 1x1).}
\label{tab:weighted_BD}
\begin{tabular}{cccccccccc||c}
\hline
\multicolumn{2}{c}{} & Market & Street & Mountain & Church & Seashore & Park & Jacuzzi & Cafe & Average \\ \hline
\multicolumn{1}{c|}{\multirow{4}{*}{$\lambda=0.1$}} & \multicolumn{1}{c|}{T. 2x2} & -27.73 & -24.53 & -19.46 & -30.02 & -15.20 & -27.82 & -29.46 & -24.82 & -24.88 \\ \cline{2-11} 
\multicolumn{1}{c|}{} & \multicolumn{1}{c|}{T. 7x7} & \textbf{-46.16} & \textbf{-45.84} & \textbf{-44.55} & \textbf{-45.84} & \textbf{-42.52} & \textbf{-45.65} & \textbf{-42.65} & \textbf{-45.16} & \textbf{-44.80} \\ \cline{2-11} 
\multicolumn{1}{c|}{} & \multicolumn{1}{c|}{ES} & 362.05 & 360.38 & 358.54 & 359.84 & 359.48 & 361.02 & 359.02 & 360.72 & 360.13 \\ \cline{2-11} 
\multicolumn{1}{c|}{} & \multicolumn{1}{c|}{RA-LDPC} & -29.05 & -28.76 & -27.39 & -28.29 & -25.82 & -28.08 & -26.42 & -27.55 & -27.67 \\ \hline \hline
\multicolumn{1}{c|}{\multirow{4}{*}{$\lambda=0.01$}} & \multicolumn{1}{c|}{T. 2x2} & -41.87 & -37.03 & -29.37 & -45.32 & -22.95 & -42.00 & -44.48 & -37.48 & -37.56 \\ \cline{2-11} 
\multicolumn{1}{c|}{} & \multicolumn{1}{c|}{T. 7x7} & -69.71 & -69.23 & -67.26 & -69.23 & -64.21 & -68.93 & -64.41 & -68.22 & -67.65 \\ \cline{2-11} 
\multicolumn{1}{c|}{} & \multicolumn{1}{c|}{ES} & -19.47 & -18.93 & -17.53 & -18.62 & -15.85 & -18.59 & -16.68 & -18.04 & -17.96 \\ \cline{2-11} 
\multicolumn{1}{c|}{} & \multicolumn{1}{c|}{RA-LDPC} & \textbf{-73.35} & \textbf{-72.11} & \textbf{-69.44} & \textbf{-71.47} & \textbf{-66.90} & \textbf{-71.69} & \textbf{-68.18} & \textbf{-70.70} & \textbf{-70.48} \\ \hline \hline
\multicolumn{1}{c|}{\multirow{4}{*}{$\lambda=1e^{-3}$}} & \multicolumn{1}{c|}{T. 2x2} & -44.11 & -39.02 & -30.95 & -47.75 & -24.18 & -44.26 & -46.87 & -39.50 & -39.58 \\ \cline{2-11} 
\multicolumn{1}{c|}{} & \multicolumn{1}{c|}{T. 7x7} & -73.46 & -72.95 & -70.87 & -72.95 & -67.67 & -72.64 & -67.87 & -71.89 & -71.29 \\ \cline{2-11} 
\multicolumn{1}{c|}{} & \multicolumn{1}{c|}{ES} & -80.15 & \textbf{-79.26} & \textbf{-77.34} & \textbf{-78.82} & \textbf{-75.56} & \textbf{-78.96} & \textbf{-76.43} & \textbf{-78.28} & \textbf{-78.10} \\ \cline{2-11} 
\multicolumn{1}{c|}{} & \multicolumn{1}{c|}{RA-LDPC} & \textbf{-80.40} & -79.00 & -76.13 & -78.34 & -73.45 & -78.63 & -74.83 & -77.57 & -77.29 \\ \hline
\end{tabular}
\end{table*}

%

\section{Discussion} \label{sec:discussion}
In the experimental section above, we have proven that it is possible to encode blocks \textit{once} and \textit{take  into account the correlation with their neighbors}, and still enable \textit{flexible extraction} during user's navigation within the 360$^\circ$ image. In particular, we have demonstrated that the tile-based approach is not suited for interactive compression because of its coarse access granularity. Indeed, only tiles can be extracted. On the contrary, our coder enables a finer granularity (at the block level) while maintaining a similar storage cost. In Section~\ref{sec:vvc}, we discuss how our solution remains compatible with most of the optimization tools embedded in  recent standards.  In Section~\ref{sec:streaming}, we discuss how classical streaming methods could still work and even benefit from our proposed solution.  Indeed, our solution is complementary rather than a competitor to these advanced tools.

\subsection{Compatibility with video coding standards}\label{sec:vvc}
As stated in the last description of the new video coding standard \cite{chen2019vvc}, the improvements in VVC brings gigantic compression gains deal mostly with: (i) the partitioning (ii) the intra prediction, (iii) the transform, quantization and (iv) entropy coding.\\
\textbf{Block partitioning} consists in optimizing the size of the blocks  based on a rate-distortion criterion. A smooth region in the image is typically processed with a large block size while textured and heterogeneous regions are split into small blocks. Whereas no block optimization has been considered in this paper (neither for tile-based approaches nor for our coder), our solution could benefit from an optimized block partitioning exactly as the standard coders. In particular, the very last partitioning \cite{chen2019vvc} that enables diagonal or non-equal block division can be implemented similarly in our approach. In particular, all these advanced tools could be adapted to our method by replacing the classical rate with our storage and transmission rate.\\
\textbf{Intra prediction} has been intensively optimized in  recent years, either by increasing the precision of the prediction (\textit{e.g.,} increasing the number of modes \cite{chen2019vvc}) or by reducing the complexity of the mode search (\textit{e.g.,} most probable mode \cite{tariq2015efficient} or rough mode decision \cite{saurty2015modified}). All these approaches can be implemented directly in our coder.\\
\textbf{Transform and quantization} has  also been optimized in  recent video standards (\emph{e.g.,} including the choice of new transforms, dependent quantization parameters (QPs) \cite{chen2019vvc}). Even though most of the new tools can be introduced in our coder, it is important to note an important difference between our solution and classical coders. While in classical coders, the signal that is transformed and quantized is the residue (\textit{i.e.}, difference between the prediction and the true block), the transformation and quantization operations of our coder are applied on the original block and the prediction themselves to avoid propagation of quantization errors (see \cref{subsec:incrementalCoder}).\\
\textbf{Entropy coding} is definitely the major difference between our coder and the standard. Indeed, the traditional arithmetic coder implemented in video standards is replaced by a rate-adaptive channel coder. Therefore, all the improvements brought to the arithmetic coder are not  directly transferable to our solution. However, a solution similar context-based adaptive binary arithmetic coding (CABAC) \cite{marpe_context-based_2003} that takes into account the context in the the compression procedure exists in channel codes \cite{TotoZarasoa2011b}, and can be considered in future work.

\subsection{Streaming}\label{sec:streaming}
As stated in \cite{hannuksela2019overview}, solutions for 360$^\circ$ image streaming are divided into two categories. The first one consists of storing different versions of the 360$^\circ$ content, each of them covering the entire sphere, but with heterogeneous resolutions over the directions (using for example the pyramidal representation \cite{kuzyakov2016next}). In other words, the server stores different orientations requested by a user and transmits the one that is closest to the requested view. The low-resolution portion of the sphere is used in case the head movement is too fast.
The second approach consists in partitioning the 360$^\circ$ picture in different tiles, each of them being coded multiple times at different bitrates, as it was done for classical 2D videos \cite{le2016tiled}. For example, a quality emphasized region approach is used to stream the proper tile at a proper quality given a user's head movement and the available bandwidth \cite{yen2019streaming,7996611}. \\
Even though the streaming of our coded bit-stream has not been tackled in this paper, the proposed coder is fully compatible with both approaches. Indeed, the blocks can be stored at different resolutions or different QPs. Our coder could even improve the aforementioned methods since, instead of storing different versions independently, one could take into account their correlation by simply adding the change of resolution/QP as new side information. For that, we can benefit from saliency-based prediction methods such as \cite{8931644} to determine important viewports and store the blocks in these regions at higher bitrates, and the blocks of less important regions at lower bitrates.


\section{Conclusion}
We proposed a new coding scheme for the interactive compression of 360-degree images with incremental rate-adaptive channel codes. Experimental results show that our scheme balances the trade-off between transmission rate and storage. In addition, the usefulness of the transmitted blocks shows that the proposed scheme better adapts to the queries of the users and the transmission rate increases gradually with the duration of the request. This avoids the staircase effect in the transmission rate that occurs in tile-based approaches and makes the encoder perfectly suited for interactive transmission. 

\section{Acknowledgment}
The authors would like to thank Laurent Guillo and Olivier Le Meur for their helpful advices on various technical issues related to this paper, and the reviewers for all of their careful, constructive and insightful comments in relation to this work.

\bibliographystyle{IEEEtran}
\bibliography{myBib}
\end{document}